\documentclass[conference]{IEEEtran}

\usepackage{booktabs}
\usepackage{graphicx}
\usepackage{lscape}
\usepackage{amssymb}
\usepackage[cmex10]{amsmath}
\usepackage[caption=false,font=footnotesize]{subfig}

\begin{document}
%
\title{ICSREF: A Framework for Automated Reverse Engineering of Industrial Control Systems Binaries}

\author{\IEEEauthorblockN{Anastasis Keliris}
\IEEEauthorblockA{Electrical and Computer Engineering\\NYU Tandon School of Engineering\\
anastasis.keliris@nyu.edu}
\and
\IEEEauthorblockN{Michail Maniatakos}
\IEEEauthorblockA{Center for Cyber Security\\NYU Abu Dhabi\\
michail.maniatakos@nyu.edu}
}

\IEEEoverridecommandlockouts
\makeatletter\def\@IEEEpubidpullup{6.5\baselineskip}\makeatother
\IEEEpubid{\parbox{\columnwidth}{
    Network and Distributed Systems Security (NDSS) Symposium 2019\\
    24-27 February 2019, San Diego, CA, USA\\
    ISBN 1-891562-55-X\\
    https://dx.doi.org/10.14722/ndss.2019.23271\\
    www.ndss-symposium.org
}
\hspace{\columnsep}\makebox[\columnwidth]{}}

\maketitle

\begin{abstract}
The security of Industrial Control Systems (ICS) has been attracting increased attention over the past years, following the discovery of real threats targeting industrial environments. Despite this attention, automation of the reverse engineering process of ICS binaries for programmable logic controllers remains an open problem, mainly due to the use of proprietary compilers by ICS vendors. Such automation could be a double-edged sword; on the one hand it could accelerate digital forensic investigations and incident response actions, while on the other hand it could enable dynamic generation of malicious ICS payloads. In this work, we propose a structured methodology that automates the reverse engineering process for ICS binaries taking into account their unique domain-specific characteristics. We apply this methodology to develop the modular Industrial Control Systems Reverse Engineering Framework (ICSREF), and instantiate ICSREF modules for reversing binaries compiled with CODESYS, a widely used software stack and compiler for PLCs. To evaluate our framework we create a database of samples by collecting real PLC binaries from public code repositories, as well as developing binaries in-house. Our results demonstrate that ICSREF can successfully handle diverse PLC binaries from varied industry sectors, irrespective of the programming language used. Furthermore, we deploy ICSREF on a commercial smartphone which orchestrates and launches a completely automated process-aware attack against a chemical process testbed. This example of dynamic payload generation showcases how ICSREF can enable sophisticated attacks without any prior knowledge.
\end{abstract}

\section{Introduction}\label{s:intro}

Industrial Control Systems (ICS) are systems used to control, monitor, and interconnect physical processes in industrial settings. A wide variety of sectors rely heavily on ICS for their operation; examples include the oil and gas industry, food processing, electric power systems, and water treatment and desalination facilities. As evident from these examples, ICS often control national critical infrastructure. Any disruption to their operation can have far-reaching consequences, ranging from severe financial losses to environmental disasters, and even loss of life~\cite{stouffer2014nist}.

Over the past years, threats originating from the cyber domain have become potential sources of disruption for ICS, following the discovery of real ICS cyberattacks~\cite{peterson2013offensive}. The most prominent example is Stuxnet, an attack against a uranium enrichment facility in Iran in 2010~\cite{falliere2011w32}. Other examples include two attacks against the Ukrainian power grid, which led to partial blackouts in 2015 and 2016~\cite{lee2016analysis, dragos2017crashoverride}, and an attack against petrochemical safety systems in Saudi Arabia~\cite{johnson2017triton}. From a financial standpoint, the annual cumulative losses of cyber incidents for large ICS companies can be in the range of \$500,000 USD~\cite{kaspersky2017state}. Correspondingly, the global ICS security market is expected to grow from \$10.24 billion USD in 2017 to \$13.88 billion USD by 2022~\cite{markets2017industrial}.

The main enabler of this new class of attacks is the ongoing convergence between Operational Technology (OT) and Information Technology (IT). To reduce costs, Commercial-Off-The-Shelf (COTS) components are increasingly being used in ICS hardware and software. At the same time, ICS components are being ubiquitously interconnected for increased situational awareness, better control, and enhanced efficiency. Despite the many benefits of the IT/OT convergence, an unwanted side-effect is the exposure of ICS environments to the same cybersecurity risks that plague the IT domain. As recent ICS attacks demonstrate, attackers are taking advantage of the expanded ICS threat landscape during this transitional period, whereas defenders appear to be lagging behind.

ICS control physical processes through Programmable Logic Controllers (PLCs). Process engineers program these controllers by developing code which is compiled to a binary that controls the target PLC. An important objective, both for actors who wish to protect, as well as actors who wish to attack ICS processes, is to \textit{automatically reverse engineer PLC binaries}. On the one hand, such capability could speed up actions after an ICS cyberattack. Digital Forensics and Incident Response (DFIR) teams could leverage automated reverse engineering to understand attack objectives of PLC malware, and timely deploy countermeasures. On the other hand, ICS malware designers could leverage the same capabilities for dynamic payloads that do not require communication with a Command and Control (C2) center, enabling sophisticated attacks even against targets that reside in air-gapped networks.

To date, to the best of our knowledge, the problem of automatically reverse engineering ICS PLC binaries remains an open problem~\cite{mclaughlin2011dynamic, krotofil2013industrial}. ICS binary reverse engineering remains a tedious manual procedure carried out by domain expert reverse engineers that are extensively familiar with the systems under study. Recent academic literature on ICS security also highlights that automated reverse engineering is an important, unsolved problem; authors either make oversimplifying assumptions regarding the binaries of interest, or assume that information for the physical process is obtained utilizing other means (e.g., espionage)~\cite{garcia2017hey, abbasi2016ghost, mclaughlin2012sabot, govil2017ladder, zonouz2014detecting}. In addition, the tools, frameworks, and techniques for reverse engineering binaries (e.g., the Interactive Disassembler --- IDA Pro) in the IT world do not directly translate to the OT world \cite{mulder2012reverse}. One of the reasons is that ICS vendors typically employ proprietary and/or not well-documented compilers for generating binaries for their target PLCs; Siemens PLCs are programmed with the SIMATIC STEP 7, Allen-Bradley PLCs with the Studio 5000 Logix Designer and the majority of other ICS vendors employ the CODESYS framework.

Identifying the gap in methodological approaches, as well as the lack of frameworks and tools specifically tailored to PLC, in this paper we propose a methodology that considers the unique characteristics of PLC binaries, and introduce the \textit{Industrial Control Systems Reverse Engineering Framework} (ICSREF). ICSREF automates the reverse engineering process for ICS binaries and can provide information on the physical characteristics of a system captured in the ICS binaries controlling it, without any prior knowledge of the system.

Automated reverse engineering of PLC binaries can enable a broad range of studies. On the one hand, the semantic-rich information stemming from ICSREF analyses can be employed for enhancing the security of ICS. A defensive use case that highlights the need for fast, automated reverse engineering of PLC binaries is \textit{PLC malware analysis}. For example, immediately after the discovery of Stuxnet, there was a pressing need for reverse engineering the malicious STEP7 payload to understand its impact on the nuclear plant it targeted. The manual reverse engineering process carried out was arduous and time consuming, delaying forensic investigations. Falliere, who reverse engineered Stuxnet, subsequently created the JEB decompiler for S7 PLCs, which automates the process and expedites analysis of Siemens S7 PLC binaries~\cite{falliereJEB}. Similarly, ICSREF can be employed to analyze PLC malware that target CODESYS-enabled PLCs, assisting DFIR professionals to quickly determine the functionality and objectives of malicious PLC binaries. When analyzing PLC malware, ICSREF can also assist in \textit{PLC code authorship attribution} by identifying malicious actors through fingerprinting the code they utilize. In the same way, ICSREF function and code snippet fingerprinting can help legitimate actors prove \textit{IP copyright infringement} by their competitors. ICSREF analyses can additionally aid \textit{source code recovery} and \textit{binary code reuse} in the case of legacy PLCs for which the original source code is not available.

On the other hand, ICSREF highlights the significance of adopting proper, proven security practices in ICS environments. Automated reverse engineering can, for example, enable \textit{dynamic process-aware payload generation}, lowering the requirements for malicious actors, and allowing sophisticated attacks against air-gapped systems without requiring prior knowledge.

Our contributions can be summarized as follows:
\begin{itemize}
    \item We propose a structured methodology for reverse engineering generic PLC binaries, which captures their intricacies and unique characteristics.
    \item We develop ICSREF, a reverse engineering framework for PLC binaries, which automatically analyzes binaries created with the CODESYS platform and fully reconstructs their Control Flow Graph (CFG).
    \item We collect and consolidate PLC source code and binaries from public code repositories, creating a database of samples for further studies.
    \item We carry out an end-to-end case study of dynamic payload generation and attack deployment against a chemical process. A commercial smartphone equipped with ICSREF automatically generates and deploys the attack, without any prior knowledge of the process.
\end{itemize}

The remainder of the paper is organized as follows: In Section~\ref{s:binaryrev} we provide background information on ICS and PLC binaries, and identify their unique characteristics compared to conventional binaries. We propose a structured reverse engineering methodology specifically tailored to PLC binaries in Section~\ref{s:meth}. In Section~\ref{s:icsref} we introduce the ICSREF framework and present its technical details. Section~\ref{s:expval} presents our techniques and results for validating ICSREF correctness, including the creation of a database of real binaries collected from public code repositories, and a performance evaluation of our framework. We demonstrate an end-to-end case study of automated attack formulation and deployment from a commercial smartphone enabled by ICSREF in Section~\ref{s:cases}. We compare ICSREF with related work in Section~\ref{s:discussion_related}, and conclude the paper in Section~\ref{s:conclusion_future}.
\section{Preliminaries}\label{s:binaryrev}
In this section we provide background information on ICS environments and outline the unique characteristics of PLC binaries\footnote{In the context considered in this work, the term \textit{binary} refers to executable files, i.e., binary files that include native machine code that is executed by a CPU with orchestration by an OS loader and/or a runtime process.} compared to conventional binaries.

\subsection{Industrial Control Systems} ICS is a broad term that encompasses various types of configurations for controlling and monitoring industrial processes. Such configurations include Supervisory Control And Data Acquisition (SCADA) systems for geographically dispersed systems, Decentralized Control Systems (DCS) for large industrial processes with autonomous controllers, and Process Control Systems (PCS) for small industrial settings. ICS can be abstracted as a set of control algorithms that operate on sets of measurement values obtained from the physical environment via sensors. The results of these control algorithms are in turn used to regulate the controlled physical processes via actuators~\cite{parr1998industrial}. The process of sensing the environment, calculating control signals and updating outputs on actuators is also known as the \textit{scan cycle} (left side of Fig.~\ref{f:sw}), and is usually performed continuously and in a real-time fashion by PLCs~\cite{stouffer2014nist}. PLCs are real-time embedded systems encased in ruggedized packages to withstand the harsh industrial environments they are deployed into. They typically include one or more microprocessors, volatile and non volatile memories, analog and digital I/O subsystems, and networking capabilities~\cite{mulder2012reverse}.

\subsection{Software development of PLC programs} The International Electrotechnical Commission (IEC) defines industry standards for PLCs in IEC 61131. More specifically, the third part of the standard, IEC 61131-3, concerns software architecture and programming of PLCs, including programming languages, data types, variable attributes, etc.~\cite{international2013international}. IEC 61131-3 describes the following graphical and textual programming languages for PLCs:

\begin{itemize}
    \item Ladder Diagram (LD), graphical
    \item Structured Text (ST), textual
    \item Function Block Diagram (FBD), graphical
    \item Sequential Function Chart (SFC), graphical
    \item Instruction List (IL), textual (deprecated)
\end{itemize}

\begin{figure}[t]
	\centerline{\includegraphics[width=0.95\columnwidth]{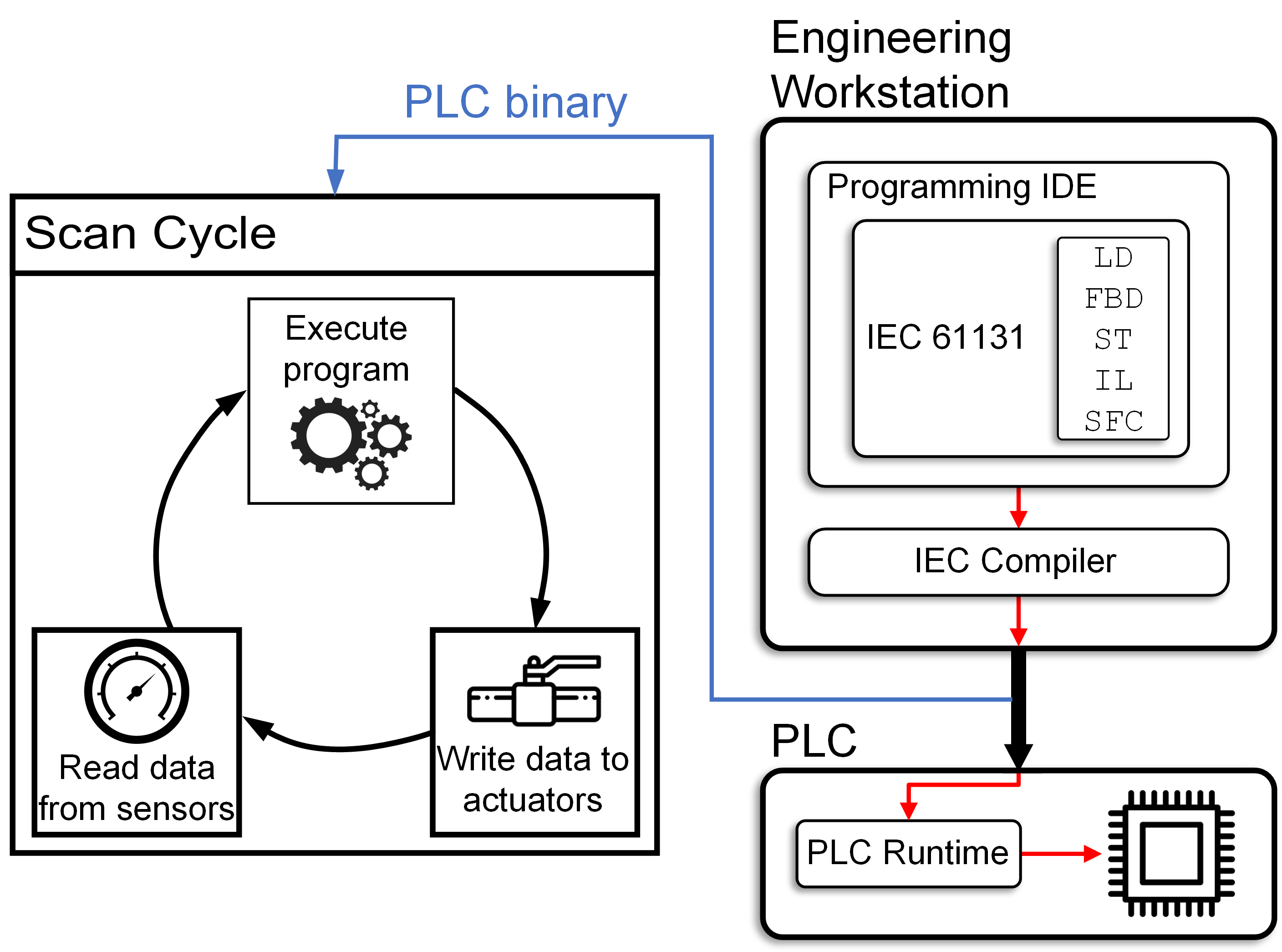}}
	\caption{Software development for PLC and scan cycle.}
	\label{f:sw}
	\vspace{-0.1in}
\end{figure}

The PLC software development process is depicted in Fig.~\ref{f:sw}. Process engineers develop PLC logic at engineering workstations. The engineering workstations are equipped with vendor-provided IEC 61131-3-compliant Integrated Development Environments (IDEs) and compilers for the specific PLC models used in the plant. PLC logic for controlling the physical process is developed using one or more of the IEC 61131-3 languages listed above and is then compiled using an IEC compiler. The binary is transferred to the PLC, a procedure called \textit{program download} in ICS terminology, where a PLC runtime (a process executed by the PLC's operating system or firmware) handles the binary loading and execution, enforcing the real-time requirements and enabling debugging and monitoring of the PLC binary execution. During normal operation the binary is loaded to the PLC's fast, volatile memory and executes from it. To ensure fast recovery in the event of an outage, the PLC binary is also stored in non-volatile memory, usually accompanied by retain/persistent variables that conserve information regarding the state of the system. From there, the PLC binary and the system execution state can be automatically loaded upon PLC startup and resume execution even after uncontrolled halts of the program.

\subsection{PLC binaries vs. conventional binaries}
Analyzing PLC binaries requires understanding their unique domain-specific characteristics that differentiate them from binaries targeting conventional workloads. Here, we highlight the main differences between the two, and discuss how these differences can facilitate or impede analyses.

\noindent\textbf{Execution model:} Besides differences in programming paradigms, execution models between conventional and PLC binaries are also different. Non-PLC languages, and by extension their compiled binaries, usually follow sequential execution of units of work (e.g., \texttt{;}-delimited statements for the C programming language~\cite{kernighan2006c}). On the contrary, the execution model of PLC binaries is dictated by the scan cycle, infinitely executing its three comprising steps. This can hinder dynamic analyses on entire PLC binaries due to their infinite execution nature, requiring that appropriate code sections are carved out for individual dynamic analyses.

\noindent\textbf{I/O operations:} Although the majority of conventional binaries rely on I/O operations to get input variables and produce corresponding outputs, the importance of I/O operations for PLC binaries is significantly higher. I/O operations in PLC binaries are a critical and necessary part of their functionality, occupying two thirds of the scan cycle. PLCs are by design devices for interacting and controlling the physical world, something that is achieved through sensor and actuator signals via the PLC I/Os. It is thus necessary to identify the mechanisms with which a PLC binary interacts and orchestrates read and write operations to these I/Os.

\noindent\textbf{File format:} Binaries compiled for major Operating Systems (OSs) usually follow well documented formats, for example the Executable and Linkable Format (ELF) for Linux and the Portable Executable (PE) format for Windows. These widely used formats are handled by loaders of their respective OSs. On the contrary, loading of PLC binaries is typically handled by proprietary loaders (e.g., the CODESYS runtime~\cite{hanssen2015programmable}), and the file format of PLC binaries is custom and unknown. This impedes the analysis of PLC binaries, as these custom proprietary formats need to be first reverse engineered to allow further exploration.

\noindent\textbf{Optimization:} Due to requirements for provable enforcement of real-time execution deadlines and assurances dictated by various standards, compilers for PLC binaries typically only make very conservative optimizations, if any~\cite{john2010iec}. On the contrary, conventional binaries that do not control critical environments typically employ several aggressive compiler optimization techniques~\cite{allen2002optimizing}. While this may lead to larger and less efficient PLC executables, it can also facilitate reverse engineering as they will be functionally simpler.

In general, these differences can be attributed to the mission-specific nature of PLCs, as well as the long history of industrial automation hardware and software that did not always follow the progression of general purpose computers~\cite{hayden2014abbreviated}. Any reverse engineering efforts should be informed by these deviations from conventional binaries.
\section{Proposed methodology for reverse engineering PLC binaries}\label{s:meth}

In this section, we propose a structured overarching methodology that is specifically tailored for reverse engineering PLC binaries. To date, several of the binary analysis approaches and techniques proposed for arbitrary binaries (e.g., function identification~\cite{bao2014byteweight} and type inference~\cite{caballero2016type}) can be leveraged for individual subtasks of PLC binary analysis. However, to the best of our knowledge, a systematic, structured methodology that covers the entire reverse engineering process for PLC binaries from start to finish is lacking. With our proposed methodology we aim to address this gap, by identifying the necessary steps and required outcomes.

When reverse engineering PLC binaries, both the similarities and differences between conventional and PLC binaries should inform the possible approaches, techniques, and tools. Taking into account their unique properties and their nature, we propose a structured methodology for enabling typical automated reverse engineering tasks and objectives, consisting of two phases: 

\begin{enumerate}
    \item A platform-specific phase, carried out once for each platform (e.g., STEP7, Studio 5000, or CODESYS). During this phase, we extract general information that characterizes and applies to all the binaries generated by that platform.
    \item An automated binary analysis phase, carried out for every binary. During the second phase, one must automate the extraction of information from a PLC binary, leveraging the platform-specific results.
\end{enumerate}

\begin{figure}[t]
	\centerline{\includegraphics[width=0.97\columnwidth]{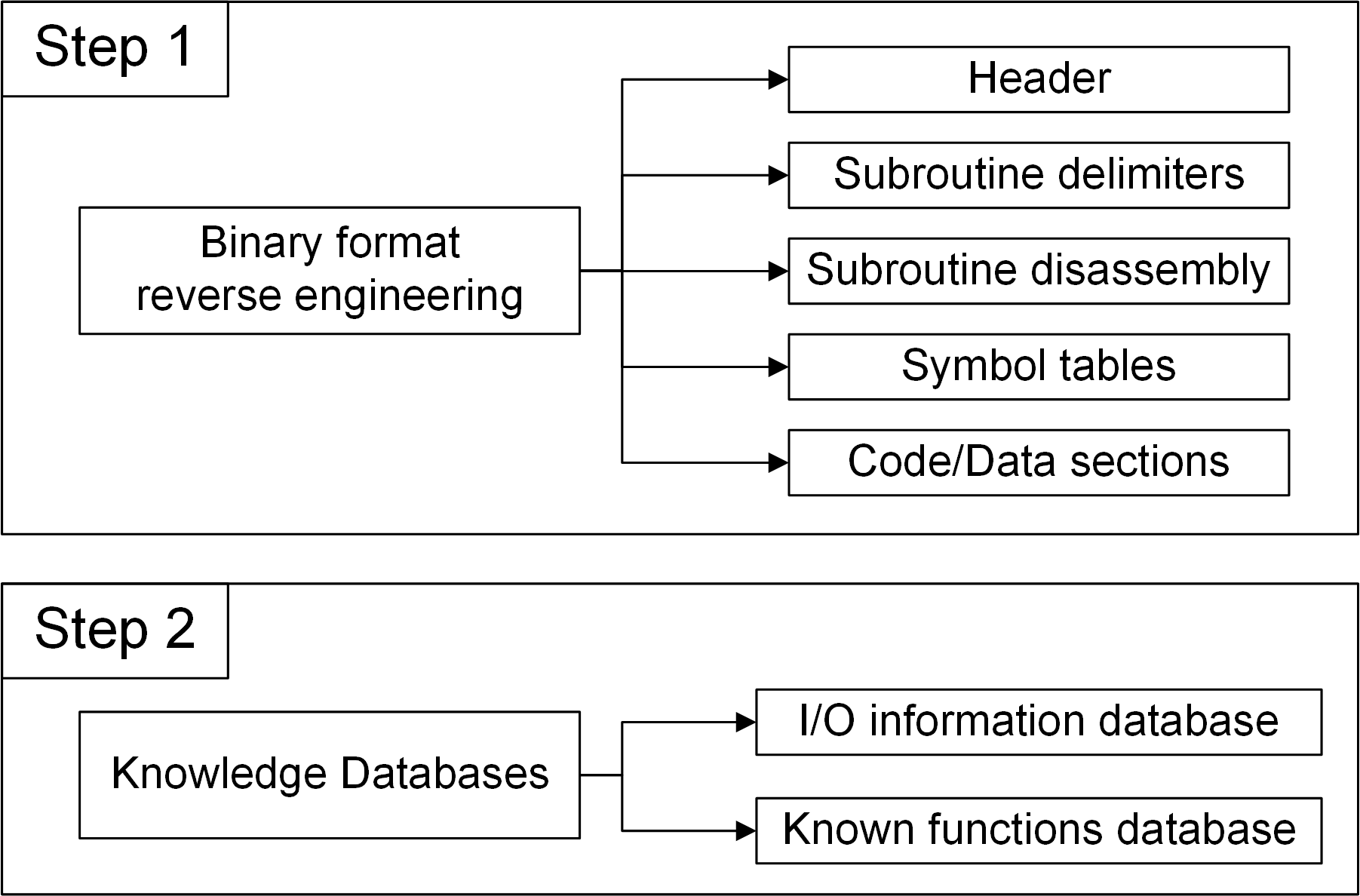}}
	\caption{Steps of platform-specific phase.}
	\label{f:offline}
	\vspace{-0.1in}
\end{figure}

\subsection{Platform-specific phase} This phase aims to extract information concerning the specifics of a platform. We break down this phase into two steps: 1) reverse engineering and understanding the general \textit{format} of PLC binaries compiled with that platform, and 2) creating \textit{knowledge databases}, that contain general information that can facilitate automated binary analyses. A schematic representation of the platform-specific phase is presented in Fig.~\ref{f:offline}. In general, this phase incurs a one-time cost per platform, and requires a minimal instantiation of the platform under analysis (i.e., a PLC device that uses this platform, and its corresponding IDE to generate PLC binaries).

\subsubsection{Binary format reverse engineering} The majority of ICS platforms employ proprietary, undocumented binary formats in their PLC binaries. Thus, the first step consists of understanding these formats and extracting relevant information. Similar to widely used executable binary formats, PLC binaries may include one or more headers, code and data sections, symbol tables for dynamically linked code, etc. There are no definitive methods for reverse engineering unknown file formats, however reverse engineering techniques from domains such as embedded systems firmware reverse engineering and file format analysis can assist in this step~\cite{cui2008tupni, skochinsky2010intro, zaddach2013embedded, eilam2011reversing}. In addition to the proprietary nature of PLC binary formats, another impediment for this step is the execution model of PLC binaries. Because of the infinite nature of the scan cycle, dynamic analyses cannot be performed on the entire PLC binary, but only on appropriately carved out instruction sequences. Nevertheless, lack of optimizations in PLC binaries may be beneficial for this step, as generalizations can be more easily made due to the immutable compilation results.

In general, the results of the binary format reverse engineering step should include information on header contents, how subroutines are delimited, extraction of the subroutines and their disassembly listings, identification of symbol tables and dynamically linked functions, as well as information on code and data sections.

\subsubsection{Knowledge Databases} For the second step, we identify the need for creating two knowledge databases containing general information that enables and accelerates reverse engineering tasks. The first should include information pertaining to I/O operations performed in the binary, given their significance for PLCs. This \textit{I/O information database} should include information on how a binary reads/writes from/to physical I/Os. In the common case of memory-mapped peripherals, the corresponding addresses for these I/O peripherals should be identified and included in the database.

The second database should contain signatures of known library functions and code snippets. These can fingerprint known subroutines in arbitrary binaries, reducing the manual effort required by a reverse engineer. This \textit{known functions database} should contain information pertaining to standard functions and libraries that can be statically linked in the binary. Similar to typical programming paradigms, IEC 61131-3 programmers can import and utilize library Functions/Function Blocks (F/FBs) in conjunction with their own F/FBs. Typical examples include F/FBs that handle network communication (e.g., MODBUS/SMTP stacks), common control algorithms (e.g., PID), and timing functions (e.g., triggers, timers). Identifying and fingerprinting these functions can speed up the analysis of a given binary, as the reverse engineer will not have to spend time understanding already known functions. A similar approach is employed by the commercial IDA Pro disassembler with its F.L.I.R.T. technology~\cite{idaflirt}. The corresponding knowledge database should employ a signature scheme that ensures low false positive and false negative rates when fingerprinting subroutines.


\subsection{Automated binary analysis} 
\begin{figure}[t]
	\centerline{\includegraphics[width=0.97\columnwidth]{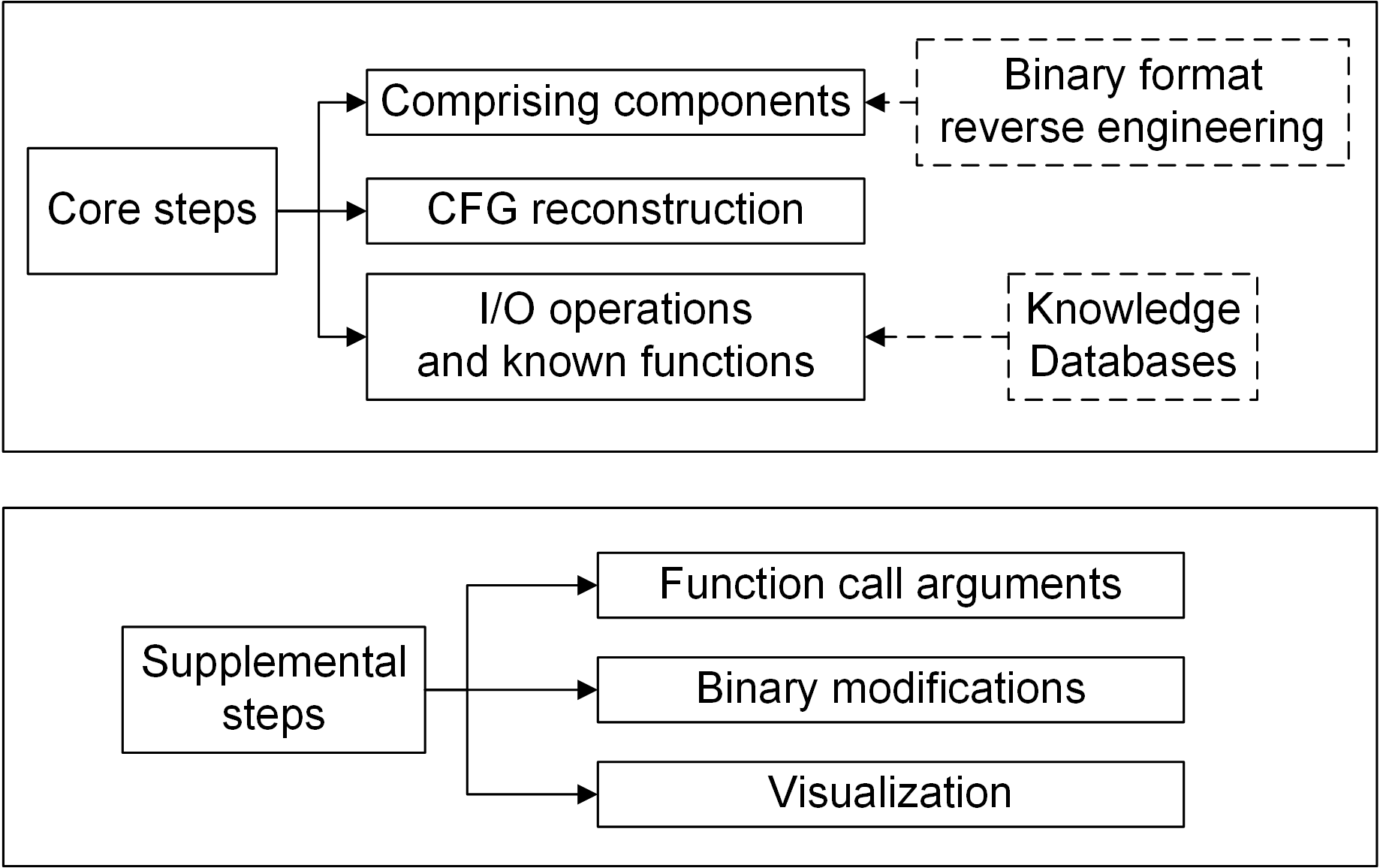}}
	\caption{Steps of automated binary analysis (Prerequisites from platform-specific phase in dotted lines).}
	\label{f:online}
	\vspace{-0.1in}
\end{figure}

The second phase automates the reverse engineering process of arbitrary binaries. It assumes that binaries are developed using a known platform analyzed during the previous phase. We identify a set of three necessary core steps for performing automated analyses, and propose supplemental steps that facilitate common reverse engineering tasks. Fig.~\ref{f:online} summarizes the requirements and components of this phase. 

The first core step should include the dissection of the binary to its comprising components, leveraging the binary file format information from the previous phase. In this step, all subroutines must be delimited and disassembled, and code/data section and symbol tables describing dynamically linked functions identified. The second core step concerns the reconstruction of an as complete and as sound as possible CFG, by finding and resolving the targets of branches. We use the definition of recovered CFG soundness and completeness from~\cite{shoshitaishvili2016state}. Finally, the third core step should identify instructions resulting to I/O operations and fingerprint known functions, utilizing the knowledge databases.

Supplemental steps can augment the resulting analysis and allow further automation. For example, a supplemental step concerns the automated extraction of arguments passed to function calls, using dynamic binary analysis techniques and symbolic execution~\cite{shoshitaishvili2016state}. This step can be leveraged to extract semantic information regarding the physical environment that a PLC binary is controlling, as this is captured by the binary. It can also assist in recovering lost source code, re-engineering and reuse of code and intellectual property audits~\cite{elwazeer2013scalable, younis2006uml}. Another supplemental step considers modification of binaries. This enables dynamic payload generation~\cite{mclaughlin2011dynamic}, or injection of host-based defenses~\cite{cui2011defending}. Binary modification should also take into account any file integrity mechanisms, such as CRC checks~\cite{Stigge06reversingcrc}. A further step should provide intuitive representation of results, including CFG visualization. In our implementation of ICSREF we implement all these supplemental steps, as described in the following section.

The structured methodology we describe in this section is general, as it is designed taking into account vendor-independent characteristics of PLC binaries. As such, it can be followed for reverse engineering of PLC binaries irrespective of development platform.

\section{The ICSREF framework}\label{s:icsref}

In this section we present the technical details of the Industrial Control Systems Reverse Engineering Framework (ICSREF), our novel modular automated reverse engineering framework. Applying our proposed methodology, we instantiate ICSREF modules that can carry out automated analyses for CODESYS-compiled binaries.\footnote{Subsequent ICSREF releases will support additional platforms.}

CODESYS is a hardware-independent IEC 61131 platform for industrial automation applications. We focus on CODESYS because it is widely used; more than 250 manufacturers employ it in their products, including multinational corporations with substantial ICS market share~\cite{hanssen2015programmable}. The CODESYS Device Directory lists 349 products or family of products that are programmed with CODESYS and are currently supported~\cite{codesys20183s}. The actual number of CODESYS-enabled devices is much larger as several products no longer appear in the directory but are still deployed. Table~\ref{t:market} lists software development platforms used by major automation corporations, demonstrating the widespread adoption of CODESYS. Due to the diverse categorization of industries involved in automation (plant automation, transportation, energy, process automation, building automation) and the sparsity of published data regarding market share of these companies, we do not have definitive data regarding CODESYS market share. However, from our domain expertise and discussions with experts, our conservative estimate is that at least 20\% of PLCs worldwide employ CODESYS.

\begin{table}
  \centering
  \caption{Automation platforms of ICS vendors}
  \label{t:market}
  \begin{tabular}{ccc}
    \toprule
    Company & Development platform & CODESYS-based?\\
    \midrule
    Rockwell Automation & Studio 5000 Logix Designer & No\\
    Siemens & STEP7 & No\\
    ABB & Automation Builder & Yes\\
    Schneider Electric & SoMachine & Yes\\
    Bosch Rexroth & Indralogic & Yes\\
    Wago Kontakttechnik & WAGO-I/O-PRO & Yes\\
    Eaton Industries & XSOFT-CODESYS & Yes\\
    Beckhoff Automation & TwinCAT & Yes\\
    Lenze Automation & PLC Designer & Yes\\
    Owen & CODESYS & Yes\\
    Omron & CX-One & No\\
    SEL & acSELerator & Yes\\
    ifm electronic & CODESYS & Yes\\
    STW Technic & CODESYS & Yes\\
    Berghof Automation & CODESYS & Yes\\
    \bottomrule
  \end{tabular}
  \vspace{-0.1in}
\end{table}

\subsection{Platform-specific phase: CODESYS}

For the needs of the first phase, we begin by reverse engineering known binaries towards understanding the binary format of CODESYS PLC binaries. Subsequently, we leverage our findings to automatically create the two knowledge databases containing PLC I/O memory maps, and known library function signatures. We carry out these studies utilizing a WAGO 750-881. The 750-881 employs an ARM microprocessor and uses CODESYS v2.3. Regardless, we stress that the results presented hereafter are not limited to binaries targeting WAGO PLCs, but are applicable to any binary compiled with CODESYS, \textit{irrespective of manufacturer}.

\subsubsection{Binary format reverse engineering} 
\begin{figure}[t]
	\centerline{\includegraphics[width=\columnwidth]{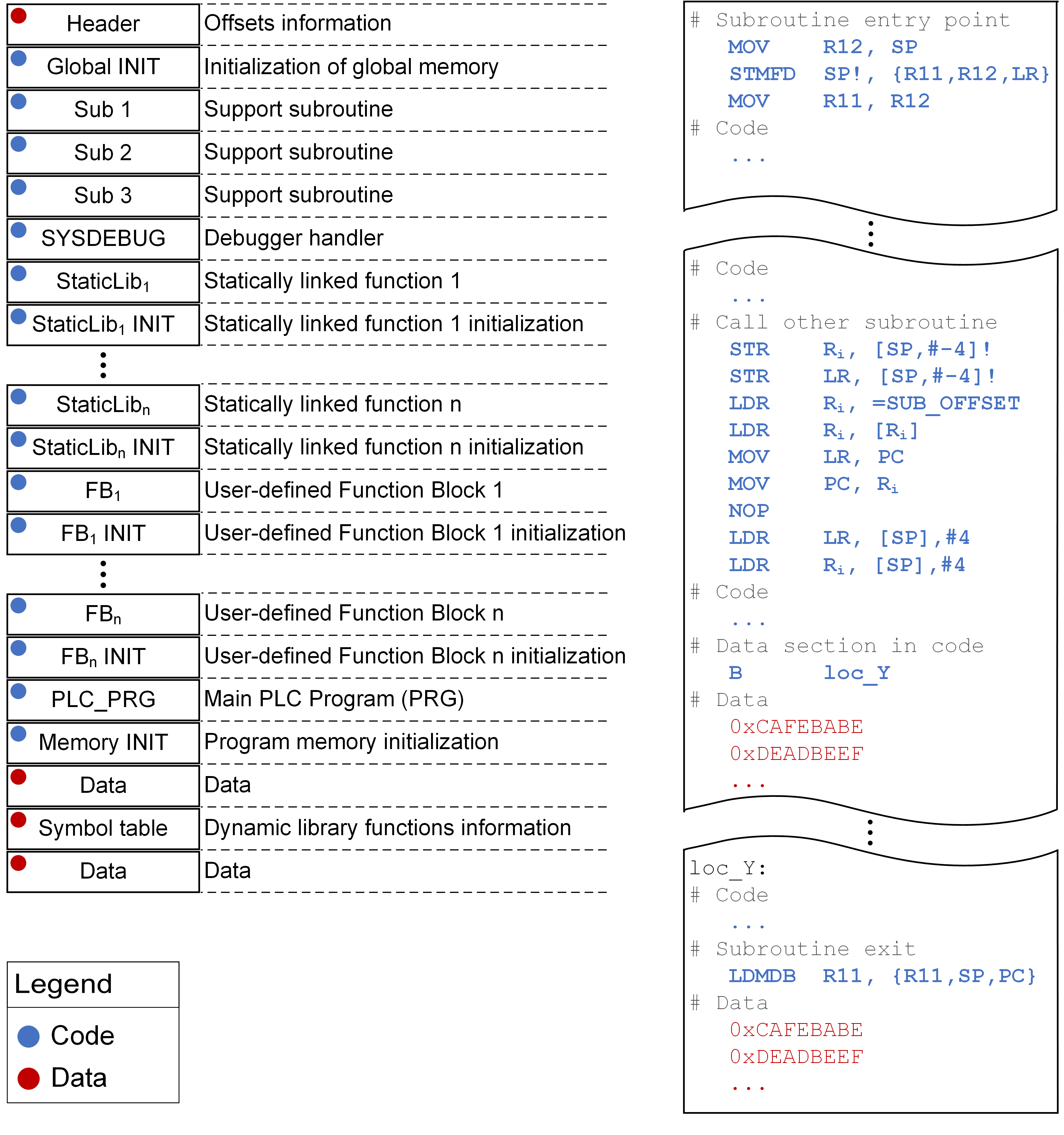}}
	\caption{Left: Format (layout) of PRG binaries, Right: Structure of subroutines, showcasing generic conventions.}
	\label{f:format}
\end{figure}

In understanding the format used in CODESYS PLC binaries, we start by developing a small number of programs that perform simple tasks. For this we utilize the CODESYS IDE, and write programs in all available IEC 61131-3 languages. We compile the programs for the ARM architecture employed by the WAGO 750-881, obtaining the corresponding PLC binaries (the file extension for WAGO CODESYS binaries is PRG). We focus on ARM binaries because of the proliferation of ARM processors in embedded systems deployed in ICS~\cite{mulder2012analysis}. Preliminary investigations of binaries compiled for other architectures indicate that the findings we describe in this section are representative of overarching CODESYS compiler mechanics, and small modifications suffice for extending ICSREF to other hardware architectures. Subsequently, we examine and reverse engineer these in-house generated PRG binaries. We disassemble the binaries and spend a considerable amount of time comparing their disassembly listings towards understanding the undocumented CODESYS binary format and the various compiler conventions. Fig.~\ref{f:format} summarizes our findings regarding binary format and code subroutines structure.

\noindent\textbf{Header:} The first 80 bytes of a PRG binary constitute a header that contains general information. Table~\ref{t:header} outlines the most important information. For example, the value obtained by adding 0x18 to the 4 byte value at offset 0x20 within the header provides the program's entry point. The resulting value is the location of the \texttt{Memory INIT} subroutine. We note that it is sufficient to only reverse engineer part of the header for enabling automated analyses.

\begin{table}
  \centering
  \caption{Header information}
  \label{t:header}
  \begin{tabular}{ccc}
    \toprule
    Offset & Length & Description\\
    \midrule
    0x04 & 4 bytes & Last global ASCII string\\
    0x20 & 4 bytes & Program entry point (-0x18)\\
    0x2C & 4 bytes & Last code subroutine (-0x18)\\
    0x30 & 4 bytes & Size of stack\\
    0x44 & 4 bytes & Last dynamic library identifier\\
  \bottomrule
\end{tabular}
\vspace{-0.1in}
\end{table}

\noindent\textbf{Subroutine delimiters:} We pinpoint the entry and exit instruction sequences (i.e., subroutine prologue and epilogue) that can delimit subroutines included in a binary. At the beginning of subroutines, CODESYS-compiled binaries use the following instructions:

\vspace{-0.3in}
\small
\begin{align*}
 & \texttt{MOV  R12, SP}\\
 & \texttt{STMFD  SP!, \{R11, R12, LR\}}\\
 & \texttt{MOV  R11, R12}
\end{align*}
\vspace{-0.2in}

\normalsize
\noindent For exiting a subroutine the following instruction is used:

\vspace{-0.2in}
\small
\begin{align*}
 & \texttt{LDMDB  R11, \{R11, SP, PC\}}
\end{align*}
\normalsize
\vspace{-0.2in}

\noindent An advantage stemming from the lack of optimizations in PLC binaries, is that these delimiters are unique and universal for all ARM CODESYS-compiled binaries. We subsequently disassemble each subroutine and study its instructions in depth to derive its functionality. Through this approach, we reverse the entire PRG binary format, as shown in Fig.~\ref{f:format}.

\noindent\textbf{Initialization of global variables and generic subroutines:} The first subroutine (\texttt{Global INIT}) starts at offset 0x50, following the header. This subroutine sets constants, variables, and initializes functions defined in the \texttt{VAR\_GLOBAL} section of an IEC 61131-3 program. It is common practice for PLC programmers to use this section for defining program-wide constants regarding the physical environment under control (e.g., scaling factors, PID gains, timing constants). After the global initialization subroutine, we observe three short support subroutines, present in every binary (\texttt{Sub 1-3}). They are followed by a debugger handler subroutine (\texttt{SYSDEBUG}) that enables extensive dynamic debugging from the IDE.

\noindent\textbf{Statically linked libraries and user-defined Function Blocks:} The debugger handler is followed by subroutines for imported library F/FBs. Each statically linked F/FB consists of two subroutines: one that performs its main functionality (\texttt{StaticLib$_i$}), and one that initializes its local memory (\texttt{StaticLib$_i$ INIT}). User-defined F/FBs that correspond to code directly developed by a PLC programmer are placed after the library F/FBs in a similar way: first a subroutine that performs their main functionality (\texttt{FB$_i$}), followed by its initialization subroutine (\texttt{FB$_i$ INIT}). We also identify the penultimate subroutine to be the \textit{main} function, a user-defined FB named \texttt{PLC\_PRG} in CODESYS. This subroutine is mandatory and serves as the starting point of the scan cycle.

\noindent\textbf{Symbol table:} Simple standard functions, such as mathematical operations on REAL-typed variables, are dynamically linked in CODESYS binaries. Information about these functions is included in a \textit{symbol table}, that is located after the last code subroutine. The symbol table contains sets of null-terminated string identifiers, followed by two bytes of data, as follows:

\vspace{-0.2in}
\small
\begin{align*}
    & \texttt{DCB "real\_add", 0} \\
    & \texttt{DCW 0x82} \\
    & \texttt{DCB "real\_sub", 0} \\
    & \texttt{DCW 0x83} 
\end{align*}
\normalsize
\vspace{-0.2in}

\noindent These two data bytes are used by the runtime to calculate the jump offset required for calling the corresponding function.



\subsubsection{CODESYS Knowledge Databases}\label{ss:db} Having understood the file format and taking into account the unique characteristics of PLC binaries, we proceed to the second step with the creation of knowledge databases. To achieve this we identify I/O operation mechanics for PRG binaries, and construct fingerprints for identifying known functions.

\begin{figure}[t]
  \centerline{\includegraphics[width=0.97\columnwidth]{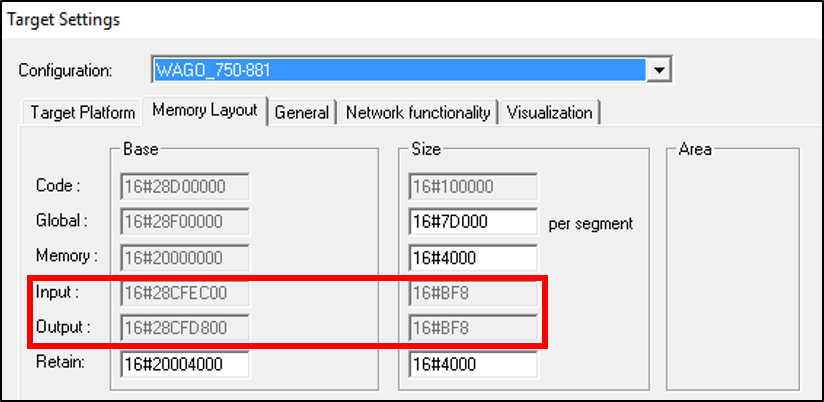}}
  \caption{I/O memory maps in CODESYS IDE.}
  \label{f:TRG}
  \vspace{-0.1in}
\end{figure}

\noindent\textbf{I/O database:} To communicate with the environment, the physical I/O modules in CODESYS-enabled PLC devices are memory mapped to specific addresses within the memory space of a PLC. These are visible in the CODESYS v2.3 IDE during creation of a new project, as seen in Fig.~\ref{f:TRG}. For example, in the case of the WAGO 750-881 PLC, whenever a memory load operation in the binary reads from a memory address within the range 0x28CFEC00 - 0x28CFF7F8 it essentially queries a sensor, and whenever a memory store operation writes in the range 0x28CFD800 - 0x28CFE3F8 it updates an actuator.

Manually extracting the values for every target PLC program from the CODESYS IDE is tedious and does not scale. By inspecting the IDE installation, we find that all architecture options for each PLC hardware are contained in target (TRG) files, residing in the installation directory of the IDE. These are data files that do not contain readable information, leading us to suspect they are encoded and/or compressed. Indeed, by further investigating the TRG file format and performing a comparative analysis of different TRG files we identify the encoding scheme used. In particular, the scheme employs exclusive disjunction (XOR) between 2048-bit blocks of the file contents and a reused 2048-bit fixed sequence. We further verified that the same scheme and sequence are used for all TRG files, irrespective of vendor and the target PLC, i.e., the technique applies to non-WAGO PLCs. This allows us to parse TRG files and extract the I/O memory maps for PLC models, populating the I/O database in an automated manner.


\noindent\textbf{Known functions database:} We create fingerprints for the statically linked library functions armed with our knowledge of the PLC binary format from the previous step. The signature scheme we select for uniquely identifying each subroutine consists of the SHA-256 hash digest of the concatenated sequence of its opcode mnemonics. We select opcodes and discard instruction arguments, because the arguments may contain mutable information (e.g., absolute branch targets, usage of different registers). As we verified in our experiments, the sequence of opcodes defines the functionality of the subroutine, i.e., the opcode sequence of a library subroutine is unique for each of our library F/FBs, and are not affected by the minimal compilation options offered by CODESYS.

\begin{figure}[t]
	\centerline{\includegraphics[width=0.97\columnwidth]{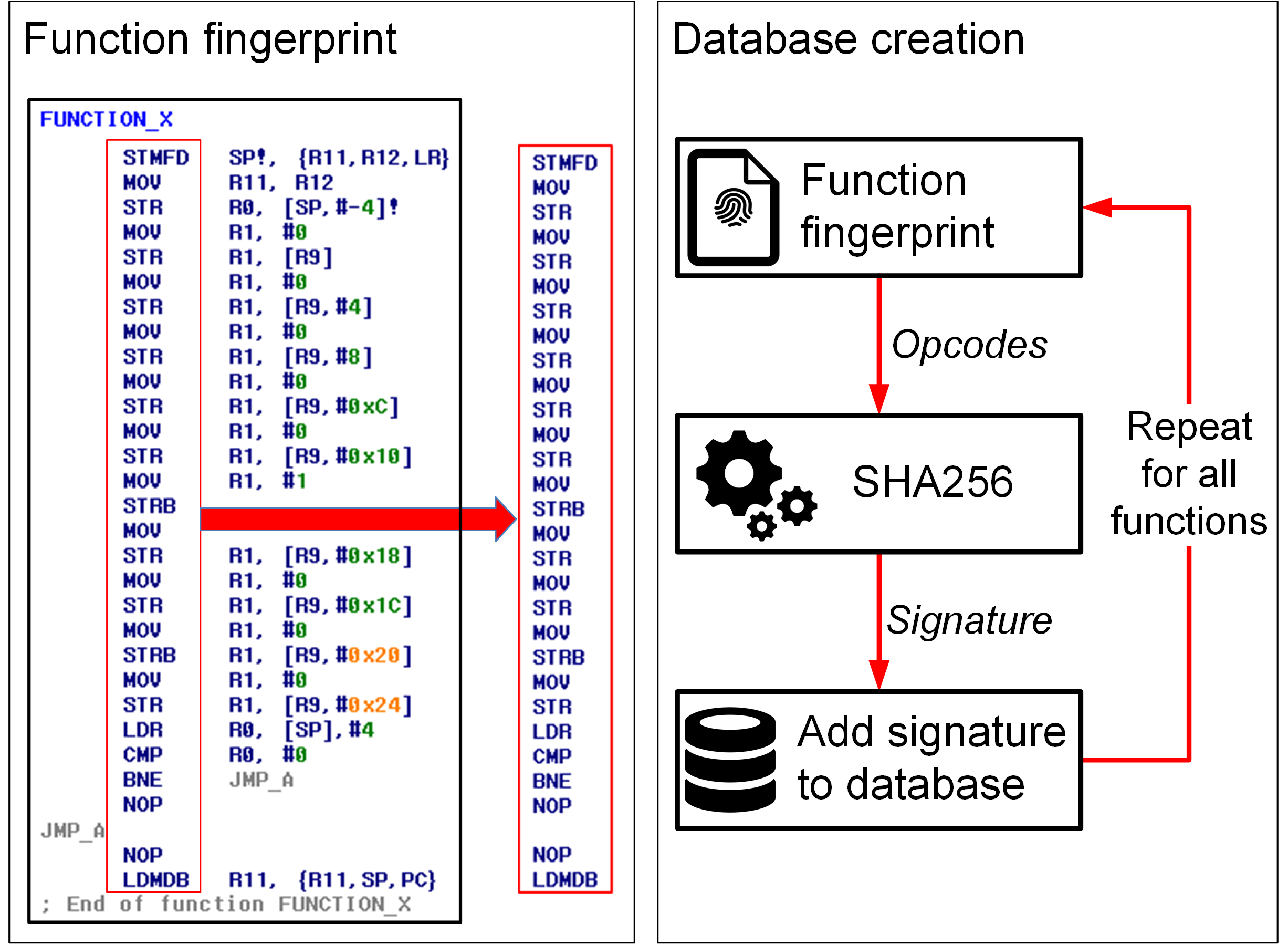}}
	\caption{Left: Fingerprint used to form function signatures, Right: Process of creating the known functions database.}
	\label{f:signatures}
	\vspace{-0.1in}
\end{figure}

We populate the known functions database with signatures for all known library F/FBs efficiently by automating all the necessary operations. Our automated database population workflow generates source code that only includes a call to one F/FB, imports the code and its requirements to the CODESYS IDE, and compiles it to obtain the corresponding PRG binary. We analyze each binary to extract the signature of the F/FB it contains, and append this signature to our database. The function fingerprint and the overall procedure for creating the known functions database is depicted in Fig.~\ref{f:signatures}.

In total, we analyze and create signatures for all 1704 F/FBs included in the 212 libraries that are available in our CODESYS IDE instantiation. The total time required for the automated process was 11 hours and 44 minutes. Although this is a time-consuming step, it is a one-time cost. After the database is created, fingerprinting subroutines in any given PRG binary resolves to a fast query of the database. During our database construction we did not find any collisions between F/FBs that perform different operations, demonstrating the suitability of using opcode sequences as identifiers. In addition, by automating this process we can easily extend the database to include F/FBs signatures from new libraries.

\subsection{Automated binary analysis: ICSREF}

We incorporate the knowledge we extract from the CODESYS platform-specific phase, as well as the automated binary analysis steps into ICSREF. ICSREF can automatically reverse engineer PRG binaries, containing modules to carry out all the steps for the second phase as they are outlined in the methodology section. Our design goals for ICSREF are:
\begin{itemize}
    \item Accuracy: The results obtained by applying ICSREF to a PLC binary should be accurate, including full CFG reconstruction whenever possible.
    \item Extensibility: ICSREF design and structure should enable easy extension to new PLC targets, microarchitectures, and IEC 61131-3 development platforms.
    \item Usability: Usage of ICSREF should be easy, and provide an intuitive interface to the reverse engineer.
\end{itemize}

To achieve these design goals we follow a modular approach, where a core module handles the core steps of the second phase. Supplemental modules extend the results, providing visualization, function call argument extraction, and binary modifications.

\subsubsection{Core module} This module processes arbitrary PRG binaries by carrying out the core steps of the automated binary analysis, namely binary dissection, CFG reconstruction, and identification of I/O operations and known functions.

\noindent\textbf{Comprising components:} To dissect a PRG binary, the core module leverages the PRG binary format information from the previous phase. First, it parses the header and extracts the information it contains. It then scans the binary searching for subroutine delimiters, and uses them to carve out all the subroutines and generate their disassembly listings using \texttt{radare2}. Finally, it extracts the symbol table and identifies any dynamically linked functions.

\noindent\textbf{CFG reconstruction:} For every binary analysis, automated or not, it is of utmost importance to reconstruct its CFG. This includes resolving the branches that are both within each subroutine, as well as jumps from one subroutine to another. A fundamental challenge that can hinder full CFG reconstruction is resolution of indirect branch targets~\cite{shoshitaishvili2016state}. Fortunately, as we explain below, PRG binaries only contain indirect jumps for calls from one subroutine to another or to a dynamically linked function, and the locations of all such indirect branch targets are completely self contained in the binary. Essentially, we can extract and resolve all indirect branches, resulting in a \textit{fully reconstructed CFG graph}. This is a byproduct of the strict requirements and standards that govern ICS platforms; lack of optimizations avoids complicated control flow operations, and provable real-time deadlines drive compilers to avoid runtime-resolved indirect jumps.

By observing the internal structure of subroutines (see right side of Fig.~\ref{f:format}), we identify the instructions that orchestrate indirect jumps for calls from one subroutine to another or to a dynamically linked function to be the following:
\begin{align*}\small
 & \texttt{STR  Ri, [SP,\#-4]!}\\
 & \texttt{STR  LR, [SP,\#-4]!}\\
 & \texttt{LDR  Ri, =SUB\_OFFSET}\\
 & \texttt{LDR  Ri, [Ri]}\\
 & \texttt{MOV  LR, PC}\\
 & \texttt{MOV  PC, Ri}\\
 & \texttt{NOP}\\
 & \texttt{LDR  LR, [SP],\#4}\\
 & \texttt{LDR  Ri, [SP],\#4}
\end{align*}\normalsize

Essentially, all jump targets (\texttt{SUB\_OFFSET}) are contained in a \textit{call table} residing in memory. In CODESYS binaries, this call table can be fully reconstructed. For dynamically linked function targets, we identify the call table index values to be calculated from the two byte data values following the null-terminated strings within the dynamic library section. Specifically, the jump offset is calculated by multiplying the two byte value of the null-terminated string identifier of a function by 4 and adding 8 to this result. For in-binary subroutine targets (e.g., statically linked library F/FBs or user-defined F/FBs), we identify that the call table is constructed by the CODESYS runtime after execution of the last subroutine, namely \texttt{Memory INIT}. The functionality of this subroutine, which is also the entry point of the binary, is twofold. First, it zero-initializes the memory space required by the binary. Second, it calculates the index offsets necessary for calling all the subroutines included in the binary, creating the corresponding call table.


We achieve extraction of the call table for in-binary subroutine targets by leveraging the dynamic symbolic (and concolic) execution capabilities of the \texttt{angr} framework~\cite{shoshitaishvili2016state,stephens2016driller,shoshitaishvili2015firmalice}.\footnote{\texttt{angr} is a python framework that combines static and dynamic symbolic analysis.} Symbolic execution allows us to execute specific parts of the binary (in this case the \texttt{Memory INIT} subroutine) without necessarily knowing the entire state of the target system, and enables extensibility of ICSREF to other platforms. An important challenge is that the loader of \texttt{angr} cannot handle PRG format binaries. We overcome this challenge by manually informing the analysis engine of the target architecture (in this case ARM) and setting a custom entry point for the binary at the beginning of the \texttt{Memory INIT} subroutine. We also force the engine to halt once it reaches the exit point of the subroutine to avoid state explosion. We then parse the memory of the resulting symbolic state, extracting all the indexes allocated for the binary's subroutine offsets, effectively reconstructing the desired call table. Following these techniques we can resolve all branch targets in any arbitrary PRG binary, effectively reconstructing a 100\% sound and complete CFG.

\noindent\textbf{I/O operations and known functions matching:} We employ the I/O operation database to identify the instructions that result to reads/writes from/to physical I/Os. For this task, we again perform symbolic execution with \texttt{angr}. In particular, we execute each subroutine individually, detecting read/write operations within the memory mapped I/Os ranges. We annotate the reverse engineering results to reflect these operations, assisting reverse engineers in identifying locations where the binary interacts with the physical environment. For matching in-binary subroutines to known library F/FBs, we calculate the signature of each subroutine, and query the known functions database for matches. Whenever there is a match, the name of the subroutine is modified in ICSREF results to reflect its functionality per its library definition.

\subsubsection{Function call arguments module} 
\begin{figure}[t]
	\centerline{\includegraphics[width=2.2in]{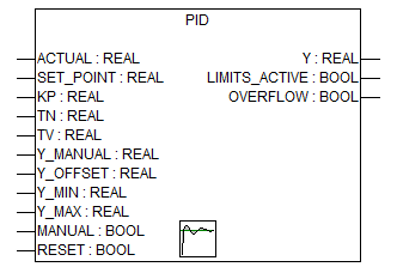}}
	\caption{PID block from CODESYS Utility library.}
	\label{f:pid}
	\vspace{-0.1in}
\end{figure}

A typical reverse engineering step in dynamic analyses concerns extracting arguments passed to subroutines at call locations. This can aid in recovering lost source code, extract semantic-based information regarding the physical environment of the PLC, and allow process-aware dynamic payload generation. We implement function call argument extraction in ICSREF by developing a module that can extract the arguments passed to Proportional-Integral-Derivative (PID) function blocks as a proof-of-concept. The PID block as defined in the CODESYS Utility library is depicted in Fig.~\ref{f:pid}. We identify that arguments are passed to a subroutine on its local stack, and reconstruct this stack using \texttt{angr}. Taking into account that the arguments could be global variables/constants, we initially symbolically execute the \texttt{Global INIT} subroutine, where these values are defined and set. Subsequently, we transfer execution to the caller subroutine and execute it until we reach a call to a function of interest (in this case PID). Finally, we query the resulting symbolic memory values on the local stack, effectively extracting the arguments that get passed to the PID function during its call. Although we develop an argument extraction module for PID functions, this module can be easily extended to extract the arguments of calls to any function for which the list of parameters and their data type are known.

\subsubsection{Binary modification module} Introducing modifications to the binary under analysis enables a plethora of studies. To that end, we develop a binary modification module for ICSREF. This module enables overwriting the contents of any location in a PLC binary (including both locations with code and data contents), with new user-specified values. Example uses include patching machine code instructions, changing branch targets, injecting executable code by overwriting ``dead code'', and modifying the call arguments of critical functions, as they are extracted with the respective module. This module can also be used to introduce malicious modifications, enabling less sophisticated actors to craft malicious PLC binaries. In the interest of protecting critical infrastructure, although we will release all ICSREF analysis modules, the binary modification module will not be disclosed publicly.

To ensure that our modified PLC binaries are considered valid by the hardware PLC, we also investigate any employed file integrity mechanisms. In particular, CODESYS uses a checksum file (CHK) to ensure the integrity of binaries. Thus, to generate valid modified binaries we need to reverse engineer this checksum algorithm. We achieve this by identifying the checksum algorithm to be an addition of all the bytes in a PRG binary, and incorporate CHK file generation in our binary modification module.

\begin{figure*}[t]
	\centerline{\includegraphics[width=0.99\textwidth]{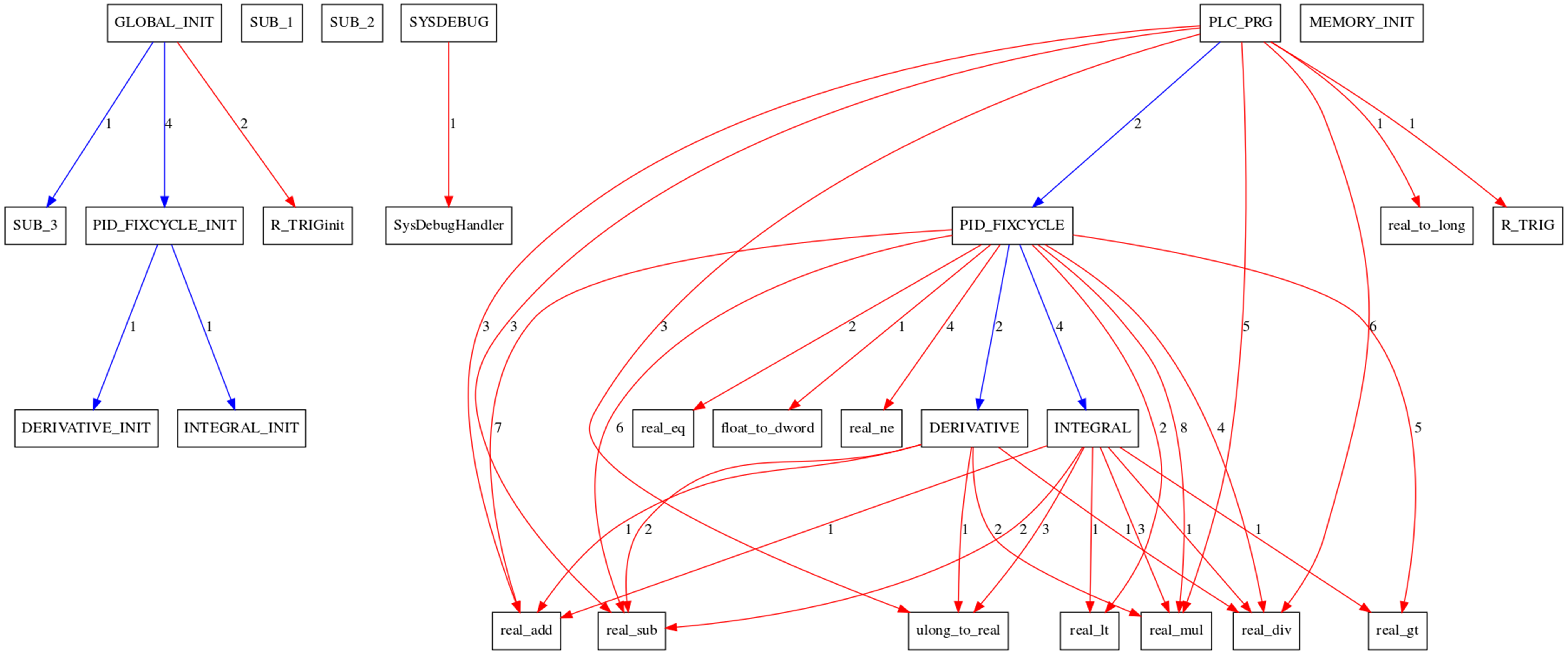}}
	\caption{Visualization of the call graph of a PLC binary that controls part of a chemical process using two PID controllers. Blue edges: Calls to statically linked functions, Red edges: Calls to dynamically linked functions, Edge weights: \# of calls}
	\label{f:graph}
	\vspace{-0.1in}
\end{figure*}

\subsubsection{Visualization module} We develop a visualization module that provides intuitive representation of ICSREF results, addressing our usability design goal. The resulting call graph visualization after applying this module to a PLC binary is shown in Fig.~\ref{f:graph}. The graph is created using the DOT graph description language, where each subroutine and dynamically linked function is a \textit{node}, and calls between them are \textit{edges}. We add weights to the edges, representing the number of calls from a caller subroutine to its callee. To differentiate between code contained in the binary and dynamically linked code, we use different colors for edges; for the former we use blue edges and for the latter red edges. Naming of unknown nodes follows the typical convention of \texttt{sub\_<OFFSET>}, whereas nodes identified by the known functions matching technique use their library definition names to assist the reverse engineer. For each node that is contained in the binary (i.e., not a dynamically linked function), we also create a hyperlink in the final SVG image that links from a node to its corresponding disassembly listing for convenient traversal.

Considering our usability design goal, we develop all ICSREF modules in Python 2.7 allowing them to execute on cross-platform systems. The reverse engineer can also directly interact with the analysis and its results from a terminal interface created using the python \texttt{cmd2} library.

\section{Experimental evaluation}\label{s:expval}

To evaluate ICSREF both in terms of correctness as well as performance we carry out extensive experiments. For our experiments we utilize in-house PLC binaries and also collect real PLC programs and binaries from public code repositories. In this section, we present the experimental results of our analyses and the details of the database of sample PLC programs we create.


\subsection{Correctness evaluation}

We verify the correctness of the platform-specific phase for CODESYS and the automated binary analysis phase by cross-checking and validating the results of analyzing a given PLC binary with ICSREF against its corresponding source code. This comparison allows a complete test since the source code that generated a binary under study is essentially a ``golden copy'' that contains all the information needed to evaluate the correctness of ICSREF results.

\subsubsection{In-house binaries}\label{ss:inhouse}

We perform a preliminary evaluation of the correctness of ICSREF using the 1704 PRG binaries, which we generated for the known functions database (see Section~\ref{ss:db}). These binaries are excellent samples for preliminary tests, as we have the source code for all 1704 binaries, and in addition, the binaries have a very rigid structure, considering that they only include a call to one library F/FB by construction. With this a priori knowledge we analyze all binaries with ICSREF, inserting automatic checks that compare the resulting CFGs with the expected CFGs stemming from the known source code structure. In all cases, ICSREF results matched the expected CFG, reconstructing a 100\% complete and sound CFG. This provides a first validation of the correctness of ICSREF results for rudimentary PRG binaries.

Subsequently, we utilize in-house developed binaries to evaluate ICSREF results when performing more complex analyses on real programs. We again compare and verify the correctness of ICSREF results against the expected functionality and CFG of the programs as it stems from their source code.

An example binary we use in this step is an IEC 61131-3 ST program targeting a WAGO 750-881 PLC. The program controls a small part of a chemical process, further described in Section~\ref{s:cases}. It contains calls to two PID control functions that execute in a fixed time interval (using \texttt{PID\_FIXCYCLE}). The program also employs a digital input that acts as a trigger using rising edge detection (using \texttt{R\_TRIG}) that orchestrates the PLC program execution using an external signal. The resulting graph of the ICSREF analysis as generated by the visualization module is presented in Fig.~\ref{f:graph}.

To verify that ICSREF results are correct, we perform a line-by-line comparison of the results against the program's source code, verifying that they match. Here, in the interest of space, we only focus on high level subroutines and their calls. As we observe from Fig.~\ref{f:graph}, the nodes and their edges as detected by ICSREF completely agree to the above description of the program operations. The main \texttt{PLC\_PRG} subroutine calls the \texttt{PID\_FIXCYCLE} function block twice, which in turn calls the \texttt{DERIVATIVE} and \texttt{INTEGRAL} functions; note that each call to \texttt{PID\_FIXCYCLE} calls \texttt{DERIVATIVE} once and \texttt{INTEGRAL} twice, something we derive by examining the source code of \texttt{PID\_FIXCYCLE} from its corresponding library file. \texttt{PLC\_PRG} also calls \texttt{R\_TRIG} for detecting the external triggering signal. The \texttt{Global INIT} subroutine initializes all the required subroutines. Several other dynamically linked function calls are made from the main function and its callees; these are either made by the imported library FBs for performing their calculations, or carry out scaling/formatting of environment values, something we verified by analyzing in depth both the developed source code, as well as all imported library F/FBs source code.

\subsubsection{Binaries from public code repositories}\label{ss:realeval}
In addition to in-house developed binaries, we collect real programs and binaries developed using CODESYS from online, public repositories on GitHub. These programs are developed by different authors, target various PLC devices, and control different, disparate physical environments. This makes them particularly appropriate for testing our framework, as their diversity can rigorously test the correctness and generality of ICSREF.

Beyond ICSREF correctness evaluation, by collecting and consolidating PLC programs and binaries we create a samples database. Our database contains both the source code as well as the corresponding binaries for industrial applications of various complexities, enabling various studies by providing a repository of diverse PLC programs. An immediate benefit of our database is that it provides benchmark programs that enable a thorough evaluation of methods and techniques that operate at the source code level, the binary level, or in between (as ICSREF does). We leverage this database for exactly this purpose in this paper. Furthermore, our database enables comparative studies with a common set of benchmark programs, and provides useful and realistic PLC program source code for educational purposes, lowering the entry bar to PLC programming and process engineering.

We construct specific search queries using the GitHub advanced search options to discover public code repositories containing CODESYS source code project files. This is not straightforward; while these files contain the source code of PLC applications, they are non human-readable files. As such, the GitHub search engine does not index them, and it cannot perform searches within their contents. We overcome this limitation by leveraging the auxiliary files that support CODESYS project files (*.pro) in order to pinpoint and collect our database samples. When developing a PLC program with CODESYS, the IDE creates several auxiliary files, such as compile information files, download/reference files, initialization and log files, and symbol files. All these files are non-text files, except the *.SYM symbol files. The symbol files comprise of ASCII characters and contain information regarding the version, project Id, checksum, and size of a project. Since they are indexed by GitHub, we use keywords present in these files (e.g., ``ProjectId''), in combination with a search for files with a SYM extension, and find the public code repositories that contain CODESYS projects.

In total, we identify and download 471 CODESYS project files from 127 repositories of 55 users across GitHub using the search terms described above. For 69 out of the 471 project files, the repositories also contained the corresponding PRG binaries. By examining the strings contained in the project files we extract the target PLC details, identifying the vendor and target PLC model. Table~\ref{t:vendors} shows the target PLC device vendors for the downloaded project files. From the table we observe that the majority of programs (68\%) are developed for WAGO PLCs. This can be attributed to two factors. First, WAGO PLCs are popular for small-to-medium installations which are also more likely to share their process engineering programs publicly. In contrast, large corporations employing devices from other vendors are arguably much more protective of their source code. As their code likely controls sensitive industrial deployments, making the code publicly available could divulge corporate secrets and intellectual property. Second, 235 out of the 471 projects we downloaded belong to a single organization, which provides software and control solutions for the energy sector based on WAGO PLCs.

\begin{table}
  \centering
  \caption{Target PLC vendors for downloaded project files}
  \label{t:vendors}
  \begin{tabular}{lcc}
    \toprule
    Vendor & Number of projects & Number of authors\\
    \midrule
    WAGO            & 320 & 19 \\
    BECKHOFF        & 71  & 16 \\
    OWEN            & 33  & 7 \\
    STW             & 24  & 1 \\
    CODESYS SoftPLC & 7   & 5 \\
    ALTUS           & 7   & 2 \\
    TTCONTROL       & 2   & 1 \\
    ifm electronic  & 2   & 1 \\
    LENZE           & 1   & 1 \\
    Googol          & 1   & 1 \\
    FESTO           & 1   & 1 \\
    Bosch Rexroth   & 1   & 1 \\
    BERGHOF         & 1   & 1 \\
    \midrule
    Total           & 471 &   \\
    \bottomrule
\end{tabular}
\vspace{-0.1in}
\end{table}

Inspection of the downloaded project files revealed that our collection includes programs in all IEC 61131-3 languages, for automation projects across several industries, including the energy sector, building automation, automotive applications, industrial deployments, robotics, etc. The diversity of these programs strengthens the power of our database, both for evaluating ICSREF, as well as for its use as benchmarks by other researchers.

For evaluating the correctness of our framework, we open the downloaded project files, import their library dependencies if we have access to them, and finally compile them for the target PLC, obtaining the corresponding PRG file. We then pass the PRG binary to ICSREF for automatic analysis. For this evaluation, our efforts are hindered by the fact that we only have access to a CODESYS IDE for WAGO PLCs, allowing us to only meaningfully compile programs developed for WAGO PLCs. Another limitation is that some of the programs require proprietary libraries that were not shared along with the program source code. Without loss of generality, for the above reasons we focus on compiled PRG binaries for WAGO PLCs, and on the 69 precompiled binaries that target PLCs from WAGO, OWEN, LENZE, and BERGHOF. After discarding the non-useful programs that are either corrupted, too simplistic, or non-functioning code stubs, we analyze 266 binaries with ICSREF.

Following the same approach as for in-house binaries, we verify the correctness of ICSREF results by comparing them against the expected functionality and CFG of the programs as it stems from their source code. Due to the complexity of the programs, their large number, and also the similarities we observed between programs developed by the same users, we rigorously compare ICSREF results versus the program source code for one program from each user. Our results indicate that similarly to the in-house binaries, ICSREF results matched the expected CFG in all cases, reconstructing a 100\% complete and sound CFG. Interestingly, during this analysis we observed that the number of subroutines contained in some binaries was less than the number of F/FBs in the corresponding source code. Our close examination revealed that this is the result of rudimentary compiler optimizations, where the compiler strips away code that is never called by the program.

\subsection{Performance evaluation}\label{s:performance}

An important requirement we set for ICSREF is that it should be capable of handling PLC binaries of arbitrary complexity and size. To assess the scalability and performance of ICSREF we measure the time required by the different steps of the analysis on our database of 266 PLC binaries.

\begin{figure*}%
    \centering
    \subfloat[radare2 time]{{\includegraphics[width=2.9in]{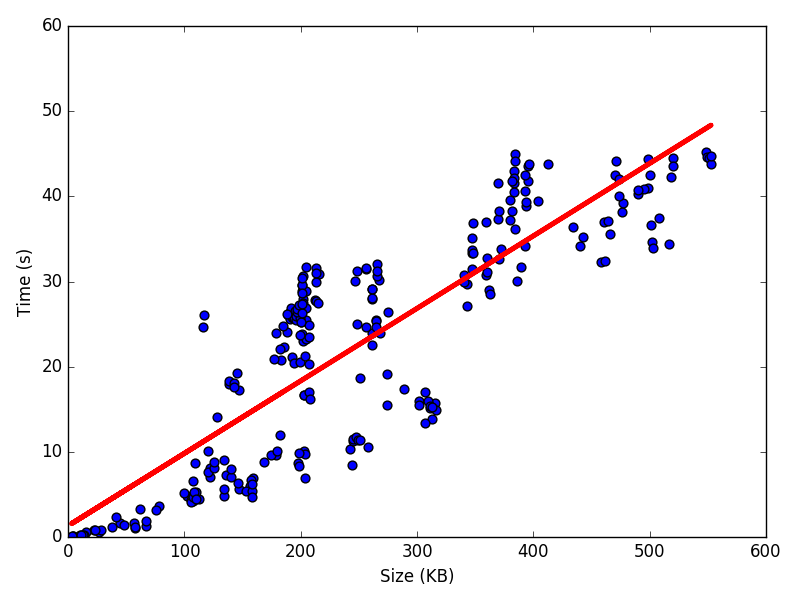} }}%
    \hspace{0.10\textwidth}
    \subfloat[angr time]{{\includegraphics[width=2.9in]{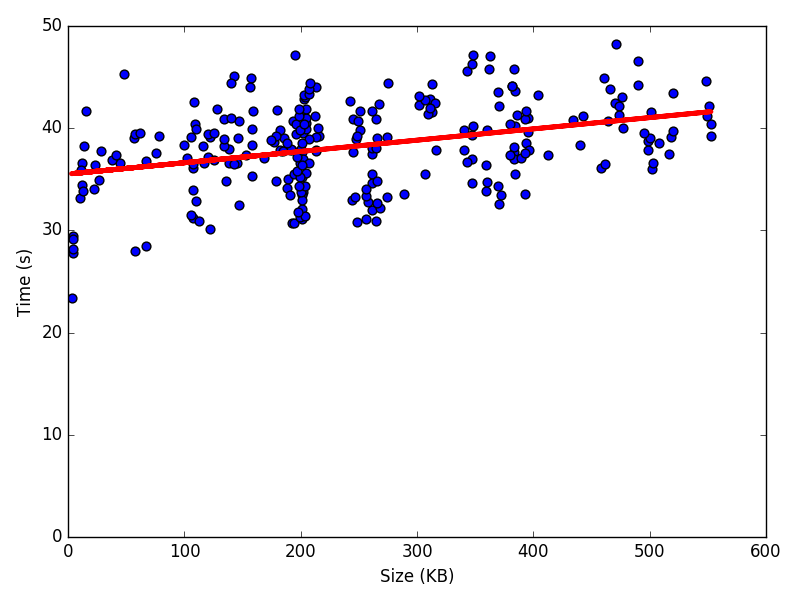} }}%
    \qquad
    \subfloat[Other operations]{{\includegraphics[width=2.9in]{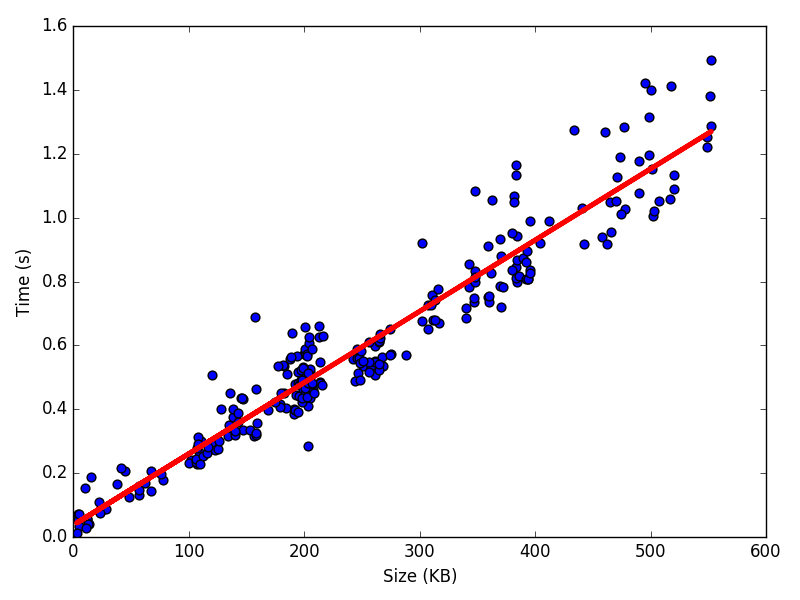} }}%
    \hspace{0.10\textwidth}
    \subfloat[Total time]{{\includegraphics[width=2.9in]{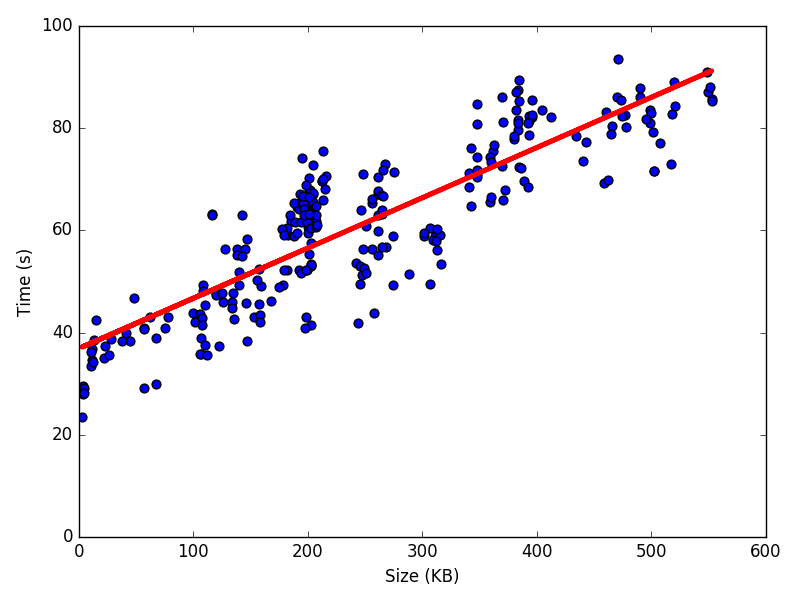} }}%
    \qquad
    \caption{Experimental results of ICSREF performance evaluation.}%
    \label{f:timing}%
    \vspace{-0.1in}
\end{figure*}

We collect the timing results for all 266 binaries by executing ICSREF on a machine equipped with an Intel i7-7500U CPU and 16 GBs of RAM, running Ubuntu 16.04. The measured times for analyzing binaries of different sizes and varying complexity are presented in Fig.~\ref{f:timing}. The graphs depict the time spent in executing the different operations of ICSREF plotted against binary size. In particular, the graphs present the time spent in: a) disassembling all binary subroutines with \texttt{radare2} in Fig.~\ref{f:timing}a, b) constructing the call table by symbolically executing the \texttt{Memory INIT} subroutine with \texttt{angr} in Fig.~\ref{f:timing}b, c) all other ICSREF operations in Fig.~\ref{f:timing}c, and finally d) the total time required to analyze a binary in Fig.~\ref{f:timing}d. From the plots in the figure, we observe that the time required by ICSREF is dominated by operations performing subroutine disassembly and call table construction. Taking an example of an average sized binary of 310 KB, its total analysis time was 58.1s, with disassembly requiring 16.0s (27.5\%), call table construction 41.4s (71.3\%) and other operations 0.7s (1.2\%).

Further analysis of the timing measurements allows us to draw conclusions on the performance of ICSREF and its consisting components. From Fig.~\ref{f:timing}a, we observe that the subroutine disassembly time has a linear relationship with the binary size. This is expected, since larger binaries contain larger and more complex subroutines whose disassembly will require more time. The time required by \texttt{angr} to symbolically execute the \texttt{Memory INIT} subroutine, depicted in Fig.~\ref{f:timing}b, is approximately constant, increasing slightly as the binary size increases. For the 266 analyzed binaries, the average time spent in this step is 38.26s with a standard deviation of 4.15s. This is also expected since the \texttt{Memory INIT} stage follows the same steps to initialize memory and construct the call table irrespective of binary size. The slight increase is attributed to the higher number of subroutines in larger programs, leading to more entries in their call tables. All other operations, depicted in Fig.~\ref{f:timing}c, require a small fraction of the overall time to execute, and intuitively have a linear relationship with binary size. The overall analysis time also has a linear dependency with binary size as shown in Fig.~\ref{f:timing}d, with a slope that is the aggregate of the slopes of its comprising operations.

The timing results demonstrate the scalability capabilities of ICSREF for handling large-size binaries. In our experience, even for complex cyber-physical systems containing hundreds of functions, the size of a PLC binary is not more than 2000 KB (note that in our database of realistic binaries no program is larger than 550 KB). Since the analysis cost is linear, ICSREF would require less than 5 minutes for processing a 2000 KB binary. This is acceptable, since the automated analysis is only performed once for every binary.

\section{Case-study: Automated attack formulation and deployment}\label{s:cases}
As mentioned above, the automated binary reverse engineering capabilities of ICSREF enable a variety of novel studies in the ICS domain. In this section, we carry out an end-to-end case study that applies ICSREF analyses in a practical scenario. In particular, we demonstrate how the automated reverse engineering capabilities of ICSREF can enable dynamic payload generation for complex ICS environments, without any prior information regarding the physical process and its controllers~\cite{mclaughlin2011dynamic}.

\noindent\textbf{Threat model:} The threat model we consider in our case study consists of two actors: an attack designer and an attack deployer. We assume the attack designer has advanced technical capabilities and designs and plans out the entire attack, but does not have physical access to the target facility. On the contrary, the attack deployer is assumed to have physical access (e.g., is a low-level employee), but lacks the technical skills for developing the attack. The deployer can be manipulated or coerced to deploy the attack. Abiding to this threat model, we consider that the attack designer bundles the attack vector in an inconspicuous everyday device, such as a smartphone, which they can hand over to the attack deployer. The deployer will in turn introduce it in the target environment, and execute the automated attack generation and deployment.

\begin{figure}[t]
	\centerline{\includegraphics[width=0.97\columnwidth]{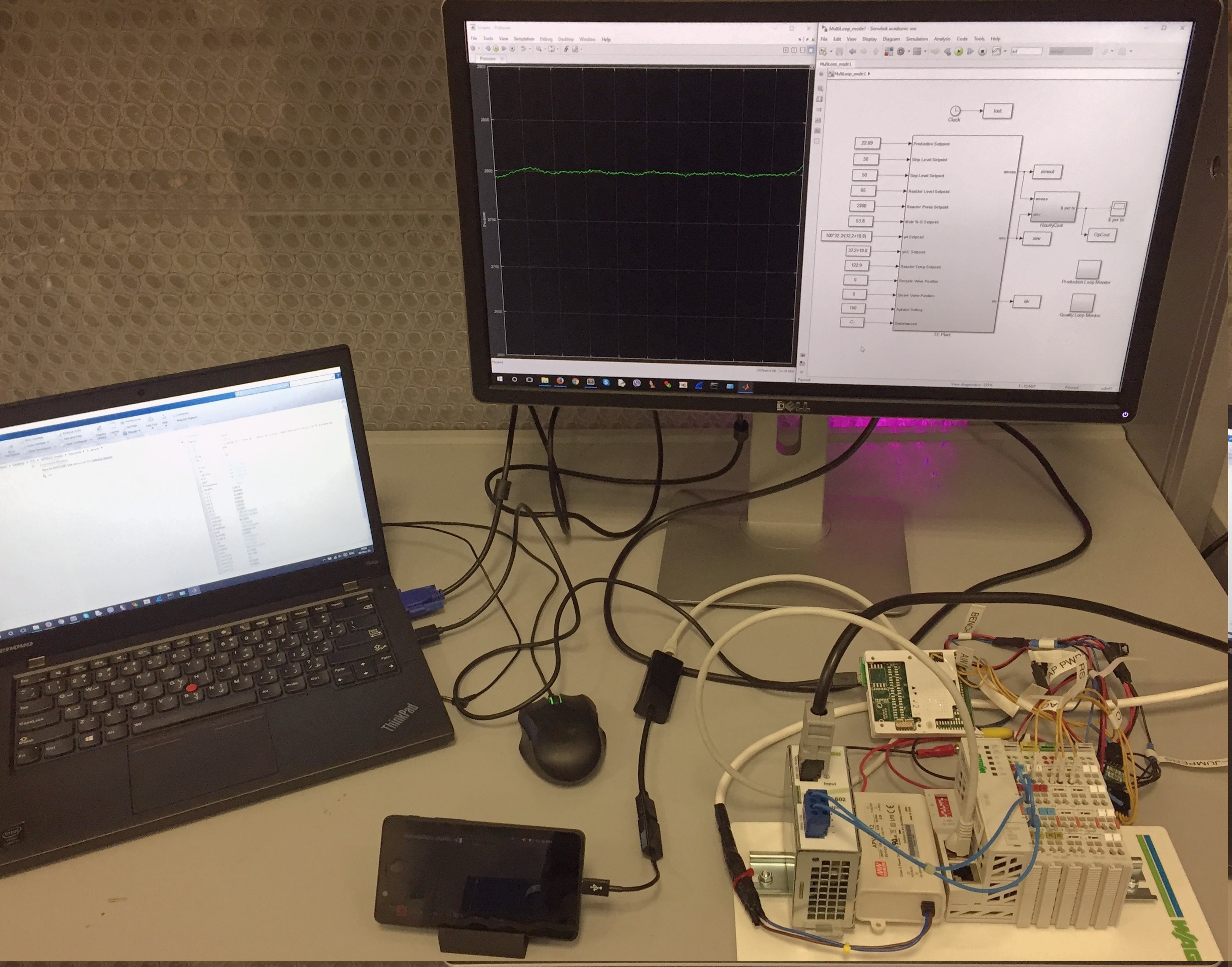}}
	\caption{Hardware-In-The-Loop testbed implementing the TE process. Top: Host PC running simulation of TE process, Bottom right: WAGO PLC connected to host PC over a serial interface, Bottom center: ICSREF-equipped smartphone for automatically generating and delivering malicious payload.}
	\label{f:testbed}
	\vspace{-0.15in}
\end{figure}

For our case study, we characterize and deploy an end-to-end automated exploitation attack vector, targeting a realistic ICS process in a controlled environment. Our target process is the Tennessee Eastman (TE) chemical process, implemented using a modified version of the TE simulation model from~\cite{tematlab}. The Tennessee Eastman challenge process is a realistic simulation of a chemical process, released to the academic community as a benchmark process~\cite{downs1993plant}. Since its release it has been widely used in several process control and ICS security studies~\cite{yin2012comparison, keliris2016machine}. In our modified version of the TE process, we incorporate real physical PLC devices that handle certain control loops of the process and communicate with the simulation environment. Specifically, we have developed a Hardware-In-The-Loop testbed, where a WAGO PLC executes two of the 18 PID controllers of the model. Fig.~\ref{f:testbed} depicts our testbed experimental setup. The functionality of the binary executing on the WAGO PLC was provided in detail in Section~\ref{ss:inhouse}.



To automatically attack the TE process, we deploy ICSREF on a commercial smartphone per our threat model, and design a simple ``push the button'' application. A smartphone is an inconspicuous device and we assume it will not raise any alarms when the deployer introduces it in the target environment. The specific smartphone we employ is an LG Google Nexus 5 running the Ubuntu Touch operating system. The attack deployer only has to establish a wired or wireless connection to the field network and launch the application. Powered by the automated reverse engineering capabilities of ICSREF, the smartphone can understand the structure of PLC binaries, make inferences regarding the physical characteristics of the controlled process, and generated modified binaries that deploy meaningful and impactful attacks in an unsupervised and autonomous manner.

\begin{figure}[t]
	\centerline{\includegraphics[width=0.95\columnwidth]{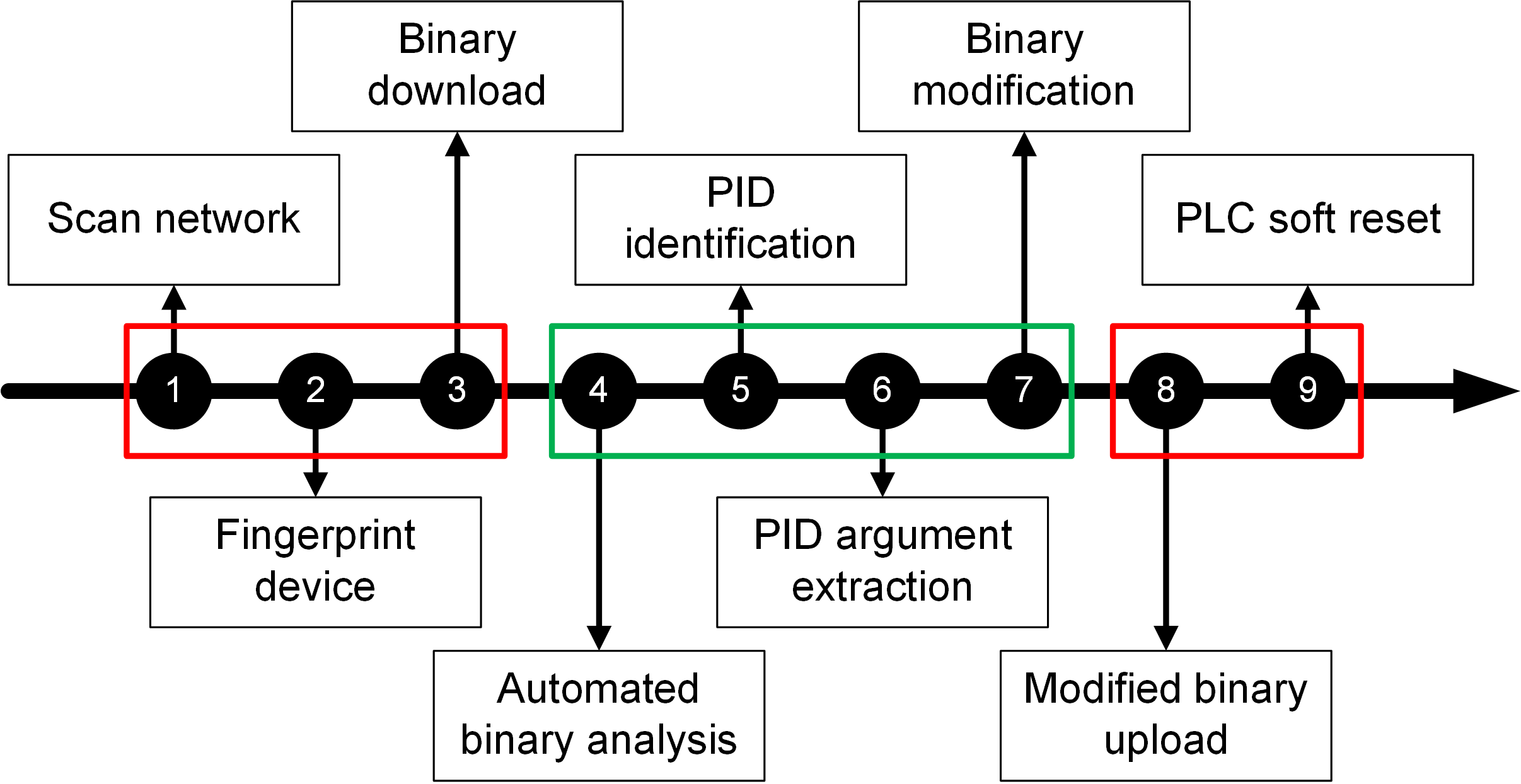}}
	\caption{Automated attack chain carried out by smartphone. Red boxes: network operations, Green box: ICSREF operations.}
	\label{f:timeline}
	\vspace{-0.1in}
\end{figure}

In our experiments, we program the phone to automatically carry out a sequence of steps, depicted in Fig.~\ref{f:timeline}. In summary, the phone scans the network for industrial devices, fingerprints them, and downloads any binaries they contain, within approximately 90s. Subsequently, the binaries are automatically analyzed with ICSREF, requiring an additional 250s. During this step, ICSREF identifies whether the binaries contain any PID functions, extracts their arguments, and changes them accordingly, generating modified binaries. Finally, the modified binaries are uploaded back to the PLC, and a soft reset command forces the PLC to execute the malicious code. This final step requires approximately 60s. Overall, the entire attack chain is executed from start to finish in less than 5 minutes. This demonstrates the feasibility of deploying context-aware payloads without orchestration in short time windows and using less powerful computing devices, such as smartphones.

\begin{figure}[t]
	\centerline{\includegraphics[width=3.0in]{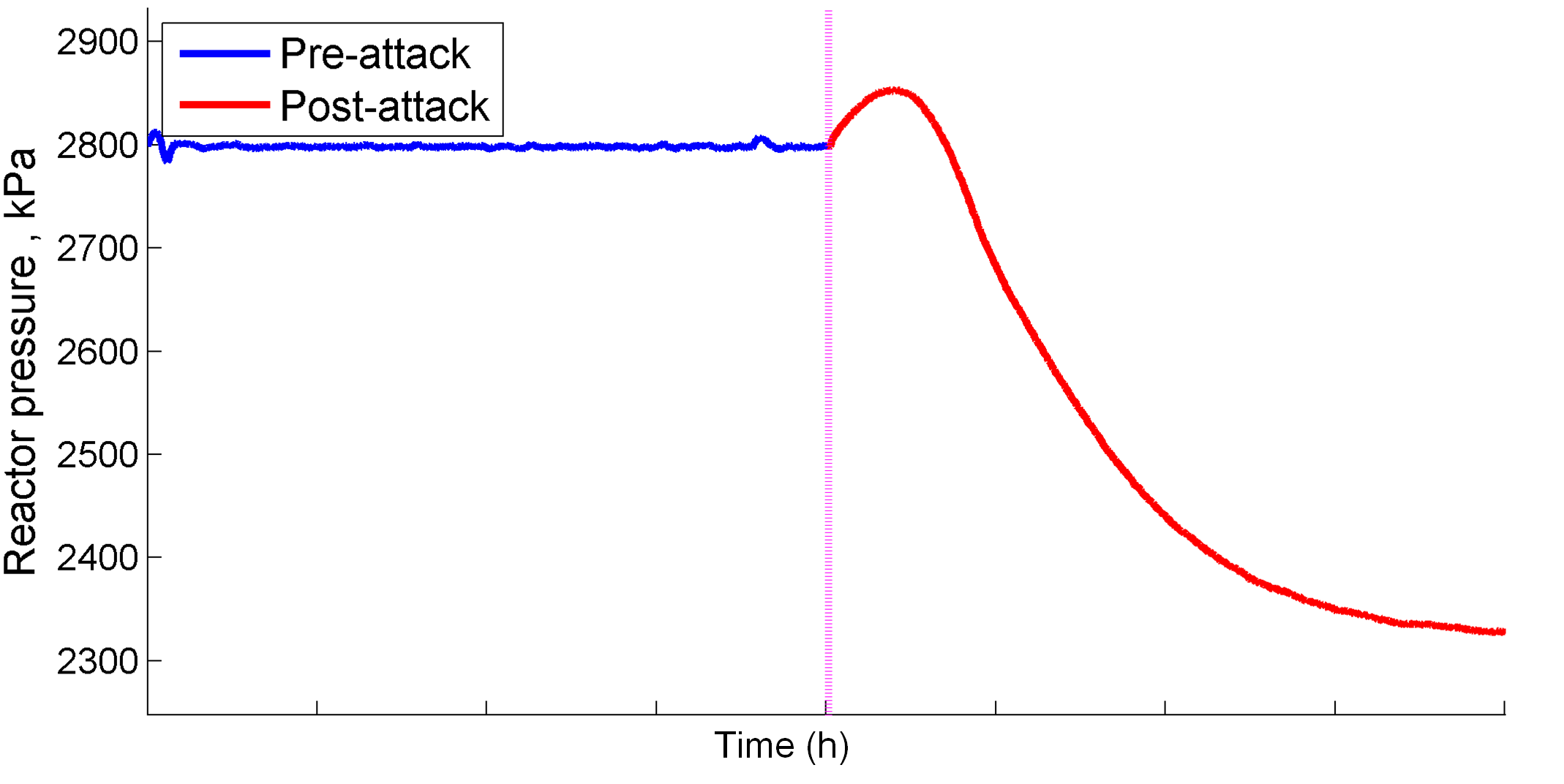}}
	\caption{Reactor pressure before and after $K_p$ payload.}
	\label{f:hitlattacks}
	\vspace{-0.1in}
\end{figure}

As a proof-of-concept we instruct the phone to generate payloads that introduce modifications that are contextually meaningful. In particular, we deploy a payload that manipulates PID functions, as these control primitives are among the most commonly used algorithms in ICS. Our ICSREF module that fingerprints known code snippets can identify PID functions, and subsequently, the function call arguments module can extract the numeric values passed to their calls. Any argument passed to the PID function during its call can be manipulated (e.g., setpoints, proportional/integral/derivative gains, initial values, etc.). Without loss of generality, we choose to alter the proportional gain ($K_p$) of the PID call that regulates the plant's reactor pressure, and increase it by a small constant value, relative to its previously programmed value (which is unknown to the phone and attacker in advance). To achieve this, we extract the current $K_p$ value with our function call arguments module, and generate a new binary that modifies it using the ICSREF binary modification module. The introduced change causes the reactor pressure to stabilize at a lower pressure setpoint (at around 2350 kPa) instead of the optimal value of 2800 kPa. The impact on the reactor pressure is shown in Fig.~\ref{f:hitlattacks}. Since this change does not cause any of the measured variables to go off range it will not trigger any alarms. However, lower reactor pressure has a negative impact on the plant-wide performance and results in monetary losses.

We stress that for automatically generating payloads without prior intelligence, it is necessary to have automated reverse engineering capabilities such as the ones provided by ICSREF. Without these, either prior knowledge regarding the binary, or interaction with a C2 server are necessary, which both render the attack more difficult and less likely to succeed. With ICSREF, operational details of the plant can be extracted from PLC binaries autonomously, and a view of the plant and its characteristics can be reconstructed on-site enabling more impactful attacks with less prerequisites.

\section{Related Work}\label{s:discussion_related}

Among the several published works concerning the security of industrial systems, the works of McLaughlin et al.~\cite{mclaughlin2014trusted, mclaughlin2012sabot} and Zonouz et al.~\cite{zonouz2014detecting} are closer to the work presented in this paper. In particular,~\cite{mclaughlin2014trusted} presents a Trusted Safety Verifier (TSV) for Allen-Bradley PLCs. TSV employs symbolic execution and model checking to verify safety-critical code that executes on a PLC. Extending this work, \cite{zonouz2014detecting} employs similar techniques to ensure that PLC code does not violate any safety requirements. SABOT, automatically generates payloads against PLCs by mapping PLC control instructions to an adversary-provided behavioral specification of the target system~\cite{mclaughlin2012sabot}. All these approaches operate on the Instruction List (IL) PLC code, which is an assembly-\textit{like} IEC 61131-3 language and not native machine code that a CPU directly executes (note that IL was deprecated in the third revision of the IEC standard). In contrast, ICSREF does not assume access to an IL version of the source code. Instead, ICSREF operates directly on the native binary, which is always present on the PLC executing it, enabling more realistic use cases. Furthermore, all 3 approaches assume full or partial knowledge of system specifics, whereas ICSREF does not require any a priori knowledge.

The proprietary, closed-source JEB decompiler for S7 PLCs by PNF Software performs similar operations to the reverse engineering analyses of ICSREF, but targets binaries for Siemens controllers~\cite{falliereJEB}. It can disassemble and decompile MC7 programs for Siemens S7 PLCs, where MC7 code is essentially proprietary bytecode interpreted by a runtime on Siemens PLCs. The disassembler/decompiler was developed after its authors manually reverse engineered the Stuxnet payload. In comparison, ICSREF targets binaries for CODESYS PLCs, providing to ability to reverse engineer additional targets. ICSREF is in contrast open source, with its modular design allowing it to be easily extended for other architectures and formats, including Siemens MC7 bytecode.

Related work also includes general purpose multiplatform decompilers, such as the state-of-the-art Hex-Rays Decompiler~\cite{hexraysdecomp} and RetDec~\cite{RETDEC_BOTCONF_17}. In general, decompilers can recreate a higher level representation of given machine code and could be useful for performing analyses of PLC binaries similar to the ones performed by ICSREF. We tested the capability of both the Hex-Rays and RetDec decompilers to analyze our database of PLC binaries and neither provided useful results. As mentioned above, this stems from the differences between IT and OT, which hinder the direct translation of IT techniques to OT scenarios~\cite{mulder2012reverse}. In contrast, the structured methodology we follow in developing ICSREF is aware of the specifics and conventions used, allowing ICSREF to successfully reverse engineer the same binaries.

Regarding ICS analyses, at a higher abstraction layer than ICSREF, operating directly on IEC 61131-3 source code, the work in~\cite{valentine2011software} parses the source code, generating a list of potential vulnerabilities and proposes associated design patterns to remove them. For attacks on PLC source code, the authors of~\cite{govil2017ladder} introduce the concept of ``ladder logic bombs'', which instantiate PLC malware written in IEC 61131-3 languages. At a lower abstraction layer than ICSREF,~\cite{basnight2013firmware} describes firmware modification attacks against PLCs, with proof-of-concept experiments against Allen-Bradley PLCs. Also at the firmware level,~\cite{garcia2017hey} presents HARVEY, a physics-aware rootkit against PLCs that can meaningfully manipulate control commands at the firmware, realizing adversary-optimal attacks. The implementation details of pin control operations in embedded systems are exploited in~\cite{abbasi2016ghost}, for instantiating stealthy attacks against a PLC and taking control of the physical process it regulates. All these works assume either full or partial prior knowledge of the target process. In contrast, ICSREF does not require any a priori knowledge of either the operations performed by a PLC binary, or of the physical environment it regulates. Nevertheless, ICSREF can be used in conjunction with these techniques to enhance their analyses, as it can automatically extract the required contextual information.

Although ICSREF is not a disassembler but a more general framework tailored to PLC binaries, a rough comparison can be made between it and more generic disassembler frameworks that also allow scripting, such as IDA Pro. We considered developing ICSREF as a set of plugins for IDA Pro, but chose to opt for an open source stand-alone framework that can be used openly and freely without requiring an IDA license. Besides its open source nature, ICSREF differs from IDA in that it is a cross platform framework. ICSREF is entirely implemented in Python for portability, and our case study in the previous section, where we deployed ICSREF on a commercial ARM-based smartphone demonstrates the cross-platform benefits of our choice. Furthermore, having full control of the development process and the selection of dependencies and software libraries in conjunction with our modular design for ICSREF, enables easy interchange of ICSREF's underlying mechanics and operation. For example, ICSREF users can easily swap \texttt{radare2} with an other disassembler, or \texttt{Graphviz} with another visualization software. In addition, the modular nature of ICSREF allows it to handle platform modifications and updates by the respective platform vendors in a robust manner. Adopting a structured methodology in implementing our framework allows us to easily capture modifications in the supported platforms and update ICSREF with the corresponding knowledge.

\section{Conclusion}\label{s:conclusion_future}
In this work we propose a structured methodology for automating the reverse engineering process of PLC binaries, taking into account the unique characteristics inherent in PLC binaries. Using this methodology, we develop ICSREF, our modular automated ICS reverse engineering framework. ICSREF can automatically analyze PLC binaries created using CODESYS, one of the three major software platforms for ICS. It can fully reconstruct the CFG of any given CODESYS binary in an automated fashion, and carry out typical reverse engineering tasks independent of vendor, PLC model, and IEC 61131-3 language. For evaluating the correctness of ICSREF we create a database of PLC programs including both source code and binaries, which can be studied and further used as benchmarks by the community. The automated capabilities of ICSREF enable a broad scope of analyses, such as dynamic payload generation which we demonstrate in our end-to-end case study. Our future work consists of enriching the capabilities of ICSREF with additional modules performing common reverse engineering tasks and extending ICSREF knowledge bases to additional platforms and hardware architectures.

\section*{Acknowledgment}

This project was supported partly by the U.S. Office of Naval Research under Award N00014-15-1-2182, and by the NYU Abu Dhabi Global PhD Fellowship program.

\section*{Resources}
ICSREF and the binary samples database can be downloaded from https://github.com/momalab/ICSREF.

\bibliographystyle{IEEEtranS}
\bibliography{icsref}

\end{document}